%% file: Final_SL5.tex
\documentclass[12pt]{article}
\pdfoutput=1
\usepackage[top=0.7in,bottom=1in,left=1.1in,right=1.1in,includehead]{geometry}
\usepackage{epsfig,amsmath,amsfonts,amssymb,verbatim}
\usepackage{float}
\usepackage{lmodern}
\usepackage[T1]{fontenc}
\usepackage{mathtext,marvosym,textcomp}
\usepackage{indentfirst}
\usepackage{epsfig,amsmath,amsfonts}
\usepackage{mathtools}
\usepackage{tabulary}
\usepackage{longtable}
\usepackage{booktabs}
\usepackage[font=footnotesize,margin=1cm]{caption}
\usepackage[usenames,dvipsnames,svgnames,table]{xcolor}
\usepackage{tikz}
\usetikzlibrary{arrows}
\usetikzlibrary{matrix}
\usetikzlibrary{shapes.misc}
\usetikzlibrary{shapes}

\tikzset{cross/.style={cross out, draw=black, minimum size=2*(#1-\pgflinewidth), inner sep=0pt, outer sep=0pt},
cross/.default={5pt}}
 \usepackage[ normalem]{ulem}
 \usepackage{tocloft}
 
 \usepackage[debug,pageanchor=false]{hyperref}
\hypersetup{colorlinks=true,linktocpage,breaklinks,
            urlcolor=blue,
            linkcolor=red,
            citecolor=blue
            }

\newcommand{\reals}{\mathbb{R}}

\numberwithin{equation}{section}

\input{Defs.tex}

\begin{document}
\renewcommand{\contentsname}{}
\renewcommand{\refname}{\begin{center}References\end{center}}
\renewcommand{\abstractname}{\begin{center}\footnotesize{\bf Abstract}\end{center}} 
 \renewcommand{\cftdot}{}

\begin{titlepage}

\vfill
\begin{flushright}
QMUL-PH-17-23

APCTP-Pre2017-018
\end{flushright}

\vfill

\begin{center}
   \baselineskip=16pt
   {\Large \bf  Exotic branes in Exceptional Field Theory: \\
   the SL(5) duality group.}
   \vskip 2cm
    Ilya Bakhmatov$^{\star \ddagger }$\footnote{\tt ilya.bakhmatov@apctp.org}, David S. Berman$^\dagger$\footnote{\tt d.s.berman@qmul.ac.uk}, 
    Axel Kleinschmidt$^{\bullet}$\footnote{\tt axel.kleinschmidt@aei.mpg.de}, \\Edvard T. Musaev$^{\bullet \ddagger}$\footnote{\tt edvard.musaev@aei.mpg.de}, Ray Otsuki$^{\dagger}$\footnote{\tt r.otsuki@qmul.ac.uk}
       \vskip .6cm
             \begin{small}
                          {\it $^{\star}$Asia Pacific Center for Theoretical Physics, Postech, Pohang 37673, Korea} \\[0.2cm]
                          {\it $^\ddagger$Institute of Physics, Kazan Federal University,
                          Kremlevskaya 16a, 420111, Kazan, Russia}\\[0.2cm]
                          {\it $^\dagger$Queen Mary University of London, Centre for Research in String Theory, \\
                          School of Physics and Astronomy, Mile End Road, London, E1 4NS, England}\\[0.2cm]
                          {\it $^\bullet$Max-Planck-Institut f\"ur Gravitationsphysik (Albert-Einstein-Institut)\\
                          Am M\"uhlenberg 1, DE-14476 Potsdam, Germany} 
\end{small}
\end{center}

\vfill 
\begin{center} 
\textbf{Abstract}
\end{center} 
\begin{quote}
We study how exotic branes, i.e. branes whose tensions are proportional to $g_s^{-\alpha}$, with $\alpha>2$, are realised in Exceptional Field Theory (EFT). The generalised torsion of the Weitzenb\"ock connection of the $\operatorname{SL}(5)$ EFT which, in the language of gauged supergravity describes the embedding tensor, is shown to classify the exotic branes whose magnetic fluxes can fit into four internal dimensions. By analysing the weight diagrams of the corresponding representations of $\operatorname{SL}(5)$ we determine the U-duality orbits relating geometric and non-geometric fluxes. As a further application of the formalism we consider the Kaluza-Klein monopole of 11D supergravity and rotate it into the exotic $6^{(3,1)}$-brane.

\end{quote} 
\vfill
\setcounter{footnote}{0}
\end{titlepage}
\setcounter{page}{2}

\tableofcontents

\section{Introduction}

In the seminal paper of Cremmer and Julia it was shown that $D=11$ supergravity compactified on a torus $\TT^n$ results in a $d=(11-n)$-dimensional theory that exhibits a hidden symmetry described by the exceptional series $E_{n(n)}$ (the split real subgroup of the complexified group) \cite{Cremmer:1978ds,Cremmer:1997ct,Cremmer:1998px}.

Subsequent to this \cite{Tseytlin:1990nb,Tseytlin:1990va,Siegel:1993th} suggested the idea that the compactified $d$--dimensional theory can be reformulated to make T-dualities manifest. Then a series of works suggested \cite{Hull:2007zu,Hillmann:2009ci,Berman:2010is} that one could make U-dualities manifest as a \emph{exceptional field theory} (EFT) living on a larger space where the U-duality symmetries appear linearly realised, i.e.\ as symmetries acting on the coordinates of the extended space that transforms under a particular representation $\mc{R}_n$ of $E_{n(n)}$. Such a space is parametrized by $\operatorname{dim}\mc{R}_n$ coordinates, $n$ of which correspond to the usual Riemannian coordinates of spacetime, whilst the other $\operatorname{dim} \mc{R}_n-n$ coordinates correspond to wrapping modes of the extended objects of M-theory, i.e.\ M2-, M5-branes, KK-monopole etc. This \emph{coordinate representation} $\mc{R}_n$ differs for each group $E_{n(n)}$ and depends on the field content of the resulting theory \cite{Berman:2011cg,Berman:2012vc}. Table \ref{reps} collects a selection of these.
\begin{table}[H]
\centering
\begin{tabulary}{\textwidth}{CCCCL}
\toprule
$n$ & $E_{n}$ & $K_{n}$ & $\mc{R}_{n}$ & Windings\\
\midrule
$3$ & $\operatorname{SL}(2)\times \operatorname{SL}(3)$ & $\operatorname{SO}(3)\times \operatorname{SO}(2)$ & ${\bf 3\oplus 2 }$ & \\
$4$ & $\operatorname{SL}(5)$ & $\operatorname{SO}(5)$ & ${\bf 10}$ & M2\\
$5$ & $\operatorname{SO}(5,5)$ & $\operatorname{SO}(5)\times SO(5)$ & ${\bf 15}$ & M2, M5\\
$6$ & $E_{6(6)}$ & $\operatorname{USp}(8)$ & ${\bf 27}$ & M2, M5\\
$7$ & $E_{7(7)}$ & $\operatorname{SU}(8)/\mathbb{Z}_2$ & ${\bf 56}$ & M2, M5, KK6\\
$8$ & $E_{8(8)}$ & $\operatorname{Spin}(16)/\mathbb{Z}_2$ & ${\bf 248}$ & M2, M5, KK6, Exotic\\
\bottomrule
\end{tabulary}
\caption{A selection of EFT symmetry groups $E_{n(n)}$, along with their maximal compact subgroups $K_n$, coordinate representation $\mc{R}_n$ and contributing windings of extended objects. The exotic objects will be discussed in more detail below.}
\label{reps}
\end{table}
\noindent A simpler case comes from compactifying $D=10$ type II supergravity on a $d$-dimensional torus, where the duality group is just the $\operatorname{O}(d,d)$ of T-duality. Then the dimension of the extended space becomes doubled $\mc{R}_{O(d,d)}=\bf{2d}$ and the extended coordinates comes from string winding modes. This is what is known as Double Field Theory (DFT) ~\cite{Hull:2009mi,Aldazabal:2013sca,Berman:2013eva,Hohm:2013bwa}.

In both DFT and EFT, the geometry of the extended space becomes closely tied to the duality group through the local symmetries of the theory and the EFT action is invariant under the so-called {\it{generalised Lie derivative}} which is described in terms of the projector $\PP$ to the adjoint representation of the duality group as
\begin{equation}
\d_\L V^M= \mL_\L V^M=\L^N \dt_N V^M +\a_n \PP^M{}_N{}^K{}_L\dt_K \L^L V^N +\l_n \dt_N \L^N V^M \, ,
\end{equation}
where the indices $M,N=1,\ldots,\operatorname{dim} \mc{R}_n$; the prefactor $\a_n$ depends on the duality group and $\l_n$ is weight of the generalised vector $V^M$. The derivative here is taken with respect to the coordinate ${\hat{Y}}^M$ on the extended space. To close the algebra of such transformations one must impose a condition on the extended space, which may be written as: 
\begin{equation}
Y^{MN}{}_{KL}\dt_{M} f\, \dt_{N} g=0. 
\end{equation}
This was first constructed for the specific case of $\operatorname{SL}(5)$ in \cite{Berman:2011cg} and for the general case of $E_{d,d}$ in \cite{Coimbra:2011ky}.
Here $f,g$ are any functions on the extended space and the $Y$-tensor $Y^{MN}{}_{KL}$ is constructed from group invariants as described in \cite{Coimbra:2011ky,Berman:2012vc} and further satisfy
\begin{equation}
\begin{split}
\label{rel}
&Y^{MN}{}_{KL}=-\a_n \PP_{K}{}^{M}{}_{L}{}^{N}+\b_n \d^M_K\d^N_L+\d^M_L\d^N_K,\\
&Y^{(MN}{}_{KL}Y^{R)L}{}_{PQ}-Y^{(MN}{}_{PQ}\d^{R)}_{K}=0, \qquad \mbox{for $d\leq5$},\\
&Y^{MP}{}_{KQ}Y^{QN}{}_{PL}=(2-\a_n)Y^{MN}{}_{KL}+(D\b_n+\a_n)\b_n\d^M_K\d^N_L+(\a_n-1)\d^M_L\d^N_K.
\end{split}
\end{equation}
Here $d=11-D$ is the number of internal directions that is being extended, $\PP_K{}^M{}_L{}^N$ is the projector onto the adjoint representation of the corresponding duality group and satisfies $\PP_K{}^M{}_P{}^Q\PP_Q{}^P{}_L{}^N=\PP_K{}^M{}_L{}^N$ and $\PP_M{}^N{}_N{}^M=\mbox{dim}(adj)$. The coefficients $\a_n$ and $\b_n$ also depend on the duality group and take on the numerical values 
$(\a_4,\b_4)=(3,\fr{1}{5})$, $(\a_5,\b_5)=(4,\fr{1}{4})$, $(\a_6,\b_6)=(6,\fr{1}{3})$.
 See~\cite{Bossard:2017aae} for a general description of tensors for any Kac--Moody group. 

In this paper we focus on the case of $\operatorname{SL}(5)$ duality group, where the coordinates transforming under the {\bf{10}} of $\operatorname{SL}(5)$ are denoted as ${\hat{Y}}^{[mn]}$ with $m,n=1,\ldots,5$ labelling the fundamental representation. To gain some intuition for the meaning of these coordinates, one decomposes them under $\operatorname{SL}(4)$ and identifies
\begin{equation}
{\hat{Y}}^{[mn]} = \begin{cases}
{\hat{Y}}^{a5} =x^a,\\
{\hat{Y}}^{ab}=\frac{1}{2} \epsilon^{abcd} y_{cd},
\end{cases} \qquad a,b=1,\cdots,4
\end{equation}
where the six coordinates $y_{cd}$ describe membrane winding modes and the four coordinates $x^a$ are the usual spacetime coordinates.
The section condition simply reads
\begin{equation}
\e^{mnklp}\dt_{mn}\otimes \dt_{kl}=0.
\end{equation}
In Section \ref{SL5EFT}, the geometry of the ${\rm SL(5)}$ theory is described in greater detail. The central object in this construction is the so-called generalised metric $\mM_{MN}$ which contains all the scalar degrees of freedom and is an element of the coset space $\operatorname{SL}(5) /\operatorname{SO}(5)$ (or more generally, of $E_{n(n)}/K_n$ where $K_n$ is the maximal compact subgroup of $E_{n(n)}$).

In the series of papers \cite{Coimbra:2011ky,Coimbra:2012af,Hohm:2013vpa,Hohm:2013uia,Hohm:2014fxa,Godazgar:2014nqa,Musaev:2014lna,Musaev:2015pla,Abzalov:2015ega,Musaev:2015ces, Berman:2015rcc} it was shown that one may consider full (supersymmetric) $d+ \operatorname{dim} \mc{R}_{n}$--dimensional theories that descend to the maximal $D=11$, half-maximal $D=10$ Type IIA/B or ungauged $d$-dimensional supergravity theories. In addition, upon a generalised Scherk-Schwarz reduction, these reproduce gauged $d$-dimensional supergravity with all gaugings expressed in terms of the generalised vielbein. The gaugings are controlled by the so called embedding tensor whose origin in EFT is the generalised torsion of the generalised Weitzenb\"ock connection \cite{Berman:2012uy,Musaev:2013rq,Berman:2013cli,Baron:2014yua}. The role of the Weitzenb\"ock connection and the generalised torsion in exceptional field theories was first discussed in \cite{Coimbra:2011ky} and later for DFT in \cite{Berman:2013uda}.

From the perspective of the lower dimensional gauged supergravity, all components of the embedding tensor may be divided into two groups: {\it{geometric}} and {\it{non-geometric}}. The former corresponds to fluxes which can be turned on when compactifying from 11 dimensions. By contrast, whilst {\it{non-geometric}} gaugings cannot be obtained via the usual compactification of 11D supergravity, they have nevertheless been shown to descend from compactifications of EFT. A good example is given in \cite{Dibitetto:2012rk} where the non-geometric gaugings of half-maximal gauged supergravity in $d=7$ were obtained from a particular choice of the generalised Scherk-Schwarz twist matrix. More information on the description of non-geometric fluxes in terms of extended geometries may be found in the works \cite{Andriot:2012an,Dibitetto:2012ia,Andriot:2012wx}. For model-building applications one may refer to \cite{Blumenhagen:2015lta}.

Just as one may consider a magnetic flux to be sourced by a monopole in electromagnetism, one may regard non-geometric fluxes as  being sourced by exotic objects. Here, by an `exotic' object we mean extended objects with tensions proportional to $g_s^{-3}$ or lower. They can be obtained by U-duality transformations of smeared conventional D-branes~\cite{Obers:1998fb,LozanoTellechea:2000mc} and have tensions proportional to $g_s^{-3}$ or lower and are thus more non-perturbative than conventional D- and NS-branes. These exotic branes indeed have also been found in string and M-theory by analysing the spectrum of massive states in $D=3$ supergravity, which fall into classifications of the $E_{8(8)}$ U-duality group \cite{deBoer:2012ma}.

For each of these exotic branes, and associated non-geometric flux, one may introduce a corresponding gauge potential of mixed symmetry\footnote{\label{fn:tensors}Throughout this paper we shall use the convention that we denote irreducible mixed-symmetry tensors of $\operatorname{GL}(D)$ by comma-separated sets of antisymmetric indices of non-increasing length e.g. \ $A_{m_1m_2m_3,n_1n_2}$ has the property of being antisymmetric in both groups of indices: $A_{m_1m_2m_3,n_1n_2} = A_{[m_1m_2m_3],n_1n_2}=A_{m_1m_2m_3,[n_1n_2]}$. Moreover, their irreducibility implies that taking the antisymmetric combination of any complete set of antisymmetric indices together with one index from the next set yields zero. In the example above, this means that $A_{[m_1m_2m_3,n_1]n_2}=0$. We shall often abbreviate the tensor by the lengths of the sets of indices, e.g. \ $A_{3,2}$ will denote a tensor of index structure $A_{m_1m_2m_3,n_1n_2}$.}, generically of the form \\ $A_{m_1\cdots m_n,n_1\cdots n_p, k_1\cdots k_q, \cdots}$,
that solves the appropriate Bianchi identity. The $D=11$ types of mixed symmetry potentials contributing to the various BPS-branes were analysed in~\cite{Bergshoeff:2011ee,Kleinschmidt:2011vu,Chatzistavrakidis:2014sua,Bergshoeff:2015cba}. For each EFT with group $E_{n(n)}$ ($n \leq 8$), one finds only a finite number of exotic branes, namely those whose fluxes can be fit into the $n$-dimensional internal space. 
A more detailed description of exotic branes and their formulation in terms of extended geometry is presented in the next section.

In terms of the conventional supergravity description, exotic branes generate backgrounds with non-trivial monodromies \cite{deBoer:2012ma}. This means that they are not globally well defined solutions in supergravity and one needs to view the supergravity as being embedded in a larger theory where the duality group is used to patch together solutions via duality transformations.

Using U-duality transformations to generate exotic branes from the conventional branes requires a sufficient number of isometries and this, in turn, implies that the brane solutions are ``smeared''. Furthermore, the brane solutions have low co-dimension and the harmonic function describing the solutions have poor behaviour at infinity (various aspects of this are discussed in \cite{deBoer:2012ma}). In a series of works \cite{Harvey:2005ab,Jensen:2011jna,Kimura:2013fda,Kimura:2013zva,Kimura:2013khz,Lust:2017jox} it was shown that including the worldsheet instanton corrections to the smeared background can localise the brane, breaking the isometry and thus recovering the full non-smeared solution, albeit with fields now depending on dual coordinates. This is already true for the Kaluza-Klein monopole whose instanton-corrected version, called the localized KK-monopole, is no longer a solution of the supergravity equations of motion. Instead, as shown in \cite{Jensen:2011jna,Berman:2014jsa}, it solves the equations of motion of DFT. In the works \cite{Bakhmatov:2016kfn,Musaev:2016yon} the T-duality chain has been investigated further and the backgrounds describing the so-called exotic Q- and R-branes have been obtained. When smeared along all the dual coordinates these correspond to the $5_2^2$ and $5_2^3$ branes of Shigemori and de Boer. Moreover, if one allows dependence on the compact winding coordinates, the Q-brane background becomes precisely the instanton corrected $5_2^2$-brane of \cite{Kimura:2013zva}.\footnote{The terminology of Q- and R-fluxes originates from the work of Wecht \cite{Wecht:2007wu} where the T-duality chain originating from a three torus with H-flux was examined and shown to produce geometric flux, $f$, and two sorts of non-geometric fluxes, labelled $Q$ and $R$. The application to branes comes from \cite{Hassler:2013wsa}.} Hence, one may consider the approach taken by DFT and EFT as one which allows us to generate instanton-corrected backgrounds of exotic branes in a systematic manner.

This papers continues research in this direction and studies the exotic branes of M-theory described in the framework of the EFT with group $E_{4(4)} \cong \operatorname{SL}(5)$. This corresponds to a $d=7$ supergravity augmented by a 10-dimensional extended space. We start with the background of the KK6-monopole, uplifted to the full (7+10)-dimensional theory, and apply an  $\operatorname{SL}(5)$ transformation to obtain the background of the $6^{(3,1)}$-brane. Previously, ~\cite{Berman:2014hna} the usual standard 1/2 BPS branes of eleven dimensional supergravity were described from an EFT perspective and shown to be given by a single EFT solution with different solutions of the {\it{section condition}}. In this paper, this description is extended to include these exotic branes showing how from the perspective of EFT, the exotic and non-exotic objects are all on the same footing. To put it in another way, exotic branes are not exotic from the point of view of EFT.

We will follow the bottom-up approach of constructing and studying the various exceptional field theories directly from the U-duality groups. In a very influential work~\cite{West:2001as} it was conjectured that $E_{11}$ should provide a fundamental theory of M-theory and subsequently the various EFTs follow from this construction~\cite{Riccioni:2007ni,West:2010rv,West:2011mm}. The particular case of ${\rm SL(5)}$ EFT was investigated in view of this in~\cite{Berman:2011jh}. Although we will not make use of the top-down construction of the ${\rm SL(5)}$ EFT, we shall sometimes use the $E_{11}$ construction to keep track of the various potentials and U-duality orbits of the solutions, in particular in section~\ref{sec:e11}. This is similar to what was done in~\cite{Kleinschmidt:2003mf,Cook:2008bi,Bergshoeff:2011ee,Kleinschmidt:2011vu,Bergshoeff:2015cba}. One should note the relevant work of \cite{Lee:2016qwn} in constructing dual fluxes in EFT and crucially the work of \cite{Blair:2014zba} in the construction of the EFT $\operatorname{SL}(5)$ fluxes.

The paper is structured as follows. In  Section \ref{exotic_intro} we give an overview of exotic branes in M-theory and their classification under a U-duality action. In this section we will use the $E_{7(7)}$ EFT which provides a concise description of the expected exotic branes. In Section \ref{orbits} the embedding tensor of the ${\rm SL(5)}$ theory is considered and it is explicitly shown how its components relate to the fluxes of the $D=(7+4)$-dimensional theory under the decomposition $\operatorname{SL}(5)\hookleftarrow \operatorname{GL}(4)$, and how one can transform the fluxes into each other by the action of an $\operatorname{SL}(5)$ transformation. Section \ref{SL5EFT}, describes the desired backgrounds whilst the corresponding fluxes are calculated in Section \ref{fluxes} to confirm the interpretation that the background is indeed sourced by the $6^{(3,1)}$-brane. Section \ref{sec:e11} we describe how the relevant U-duality transformation are described in $E_{11}$. This focuses on the transformations from the M5 to the $5^3$ and from the KK6 to the $6^{(3,1)}$ and then relates these transformations to Weyl reflections involving the exceptional node. Finally Section \ref{discussion} reflects on the nature of these solutions. In particular how non-geometry arises from winding mode dependences.

\section{Exotic branes in Type II and M-theory}
\label{exotic_intro}

We will start in subsection~\ref{exotic_intro_1} by giving a systematic account of the exotic branes with masses proportional to $g_s^{-3}$ along with their fluxes, and the M-theory objects that they are related to. This information is summarised in Figure~\ref{dia_br} and Table~\ref{tab:BranesAndPotentials} below. Note that we shall be considering the $E_{7(7)}$ EFT here only as a means to simplify the presentation. The $E_{7(7)}$ EFT is particularly convenient to describe the structured picture, compared to the $\operatorname{SL}(5)$ EFT, which is the main focus of this article.

In subsection~\ref{exotic_intro_2} we turn our attention to the $\operatorname{SL}(5)$ theory. We show which exotic branes fit into which representations of the U-duality group and how the embedding tensor decomposes. This lays the foundations for the study of the U-duality rotations in subsequent sections.

\subsection{\texorpdfstring{$g_s^{-3}$}{gs-3}-branes from EFT}
\label{exotic_intro_1}

The generalised torsion of the $E_{7(7)}$ EFT has components transforming only in the $\bf 912$ and the $\bf 56$ representations of $E_{7(7)}$~\cite{deWit:1983gs,deWit:2005hv,Riccioni:2007au,Bergshoeff:2007qi,deWit:2008ta,Hohm:2013uia}. These are decomposed under the embedding $E_{7(7)} \hookleftarrow \operatorname{SO}(6,6)\times \operatorname{SL}(2)$ as follows:
\begin{equation}
\begin{aligned}
\bf 912 &\longrightarrow \bf (12,2)\oplus(32,3)\oplus(220,2)\oplus(352,1),\\
\bf 56 &\longrightarrow \bf (12,2)\oplus(32,1).
\end{aligned}
\end{equation}
The group $\operatorname{SL}(2)$ is the one that transforms the axio-dilaton, and hence by further decomposing $\operatorname{SL}(2)\hookleftarrow \reals_+$ to obtain the dilaton scaling and thus a relation to the $g_s$ power, we obtain the following:
\begin{equation}
\begin{aligned}
\bf 912 &\longrightarrow \bf 32_{-2}\oplus 220_{-1}\oplus 12_{-1}\oplus 352_0 \oplus 32_{0}\oplus 220_{+1} \oplus 12_{+1}\oplus 32_{+2},\\
\bf 56 &\longrightarrow \bf 12_{-1}\oplus 32_0 \oplus 12_{+1}.
\end{aligned}
\end{equation}
The $\reals_+$ weight here is naturally related to the power of $g_s$ appearing in the expression for the mass for the corresponding state in 4D maximal supergravity or, equivalently, in the expression for tension of the corresponding brane (see below). The relation to this power of $g_s$ is obtained by an appropriate shift of the $\reals_+$ weight. We shall follow the conventions of \cite{Lombardo:2016swq} and identify the irreducible representation $\bf 32_2$ with the RR fluxes ($T\sim g_s^{-1}$), $\bf 220_1$ with the NS-NS fluxes ($T\sim g_s^{-2}$) and the $\bf 352_0$ with the non-geometric P-fluxes ($T\sim g_s^{-3}$). The others correspond to non-geometric fluxes sourced by branes whose tension is proportional to even lower powers of $g_s$.

To single out the fluxes in their conventional six-dimensional tensorial form, one must further decompose the representations under the embedding of $\operatorname{O}(6,6)\hookleftarrow \operatorname{GL}(6)$, that gives the fluxes for $g_s^{-\alpha}$ with $\a \leq 3$
\begin{equation}
\begin{aligned}
\bf 32_2 & \longrightarrow \bf 1_{-3/2}\oplus15_{-1/2}\oplus\overline{15}_{1/2}\oplus 1_{3/2} \\
\bf 220_1 & \longrightarrow \bf 20_{-3/2}\oplus \bar{6}_{-1/2}\oplus\overline{84}_{-1/2} \oplus84_{1/2}\oplus6_{1/2}\oplus20_{3/2} \\
\bf 352_0 & \longrightarrow \bf 35_{-3/2}\oplus15_{-1/2}\oplus21_{-1/2}\oplus105_{-1/2}\oplus\overline{105}_{1/2}\oplus\overline{21}_{1/2}\oplus \overline{15}_{1/2}\oplus35_{3/2}
\end{aligned}
\end{equation}
where the conventions are chosen to be adapted to the IIA reduction and the labels are related to the scaling of the object's tension with the M-theory circle. At the level of tensors, the above decomposition reads 
\begin{equation}
\begin{aligned}
\a=1&& \q_A & \longrightarrow  F \oplus F_{a_1a_2}\oplus F_{a_1\ldots a_4} \oplus F_{a_1\ldots a_6} \\
\a=2&& \q_{MNK} & \longrightarrow  R^{abc}\oplus f^a\oplus Q_a{}^{bc} \oplus f_{ab}{}^c\oplus f_a\oplus H_{abc} \\
\a=3&& \q_{M \dot{A}} & \longrightarrow P^a{}_b\oplus P^a{}_{b_1b_2b_3}\oplus P^a{}_{b_1\ldots b_5}\oplus P_{b_1b_2} \oplus P_{a,b}\oplus P_{a,b_1b_2b_3}\oplus P_{a,b_1\ldots b_5} \oplus P_{b_1\ldots b_4}.
\end{aligned}
\end{equation}
Note that the fluxes at the level $\a=2$ contain the components $f^a$ and $f_a$, which correspond to the gaugings $\F_M$ of half-maximal supergravity. Such gaugings do not correspond to fluxes sourced by BPS branes or trombone gaugings of the maximal theory. They are instead equal to the trace parts of the components $\F_{ab}{}^c$ and $\F_a{}^{bc}$ of the generalised torsion, with the remaining traceless components being the regular $Q$ and $f$ fluxes. The same is true for the fluxes $P_{ab}$ and $P_{abcd}$, which shall be omitted from further discussion together with $\F_M$.

For the Type IIB reduction one would obtain the RR fluxes inside the $\q_{\dot A}$ tensor and the P-fluxes inside the $\q_{M A}$ tensor (of opposite chirality). Alternatively, one could embed the $\operatorname{GL}(6)$ differently. 

In $D=10$ one generally finds all the fluxes corresponding to $\a=3$ inside the tensor $\q_{M \dot A}$ (or $\q_{M A}$) of $\operatorname{O}(10,10)$, for which we write
\begin{equation}
\begin{aligned}
\text{IIA:}&& \q_{M\dot A} \longrightarrow &\ P_{a, b}\oplus P_{a,b_1\ldots b_3}\oplus P_{a, b_1\ldots_5}\oplus P_{a, b_1\ldots b_7}\oplus P_{a, b_1\ldots b_9}  \\ 
&&  &\ \oplus P^a{}_{b}\oplus P^a{}_{b_1\ldots b_3}\oplus P^a{}_{b_1\ldots_5} \oplus P^a{}_{b_1\ldots b_7}\oplus P^a{}_{b_1\ldots b_9}, \\ \\
\text{IIB:}&& \q_{M A} \longrightarrow &\ P_{a}\oplus P_{a,b_1b_2}\oplus P_{a, b_1\ldots_4}\oplus P_{a, b_1\ldots b_6}\oplus P_{a, b_1\ldots b_8}\oplus P_{a, b_1\ldots b_{10}}  \\ 
&&  &\ \oplus P^a\oplus P^a{}_{b_1b_2}\oplus P^a{}_{b_1\ldots_4} \oplus P^a{}_{b_1\ldots b_6}\oplus P^a{}_{b_1\ldots b_8} \oplus P^a{}_{b_1\ldots b_{10}}.
\end{aligned}
\end{equation}
Here, we are using the notation common in the DFT literature and are decomposing the $\operatorname{O}(10,10)$ generalised index $M =1, \cdots, 20$ into the two sets $({}_a,{}^a)$ with $a = 1, \cdots, 10$.

Since we know how T-duality acts on the irreducible representations of $\operatorname{O}(10,10)$ it is straightforward to recover the T-duality transformation rules for the P-fluxes. Indeed, the first index of $P_{x,b_1\ldots b_n}$ (resp. $P^x{}_{b_1\ldots b_n}$), which comes from the vector index of $\operatorname{O}(10,10)$, is lowered (resp. raised) if the T-duality transformation $T_x$ acts along $x$. The rules for indices in the second set are the same as for the RR-fluxes, i.e.\ if the direction $x$ is in this set, then $T_x$ removes the corresponding index and adds it if it is not. All of this is summarized as follows \cite{Lombardo:2016swq}:
\begin{equation}
\begin{aligned}
T_a:& P_a{}^{b_1\ldots b_p}\longleftrightarrow P^{a,b_1\ldots b_p a},\\
T_{b_p}:& P_a{}^{b_1\ldots b_p} \longleftrightarrow P_a{}^{b_1\ldots b_{p-1}}.
\end{aligned}
\end{equation}

In \cite{Bergshoeff:2011ee,Lombardo:2016swq} it was shown that the corresponding gauge potentials for the $P$-fluxes listed above are the following
\begin{equation}
\begin{aligned}
\text{IIA}:&\  E_{8,1}\oplus E_{8,3}\oplus E_{8,5}\oplus E_{8,7}\oplus E_{10,3,2}  \\ 
& \ E_{9,1,1}\oplus E_{9,3,1}\oplus E_{9,5,1} \oplus E_{9,7,1}\oplus E_{10,5,2}, \\ \\
\text{IIB}:& \ E_{8,0}\oplus E_{8,2}\oplus E_{8,4}\oplus E_{8,6} \oplus E_{10,2,2}   \\ 
&  \ \oplus E_{9,2,1}\oplus E_{9,4,1} \oplus E_{9,6,1}\oplus E_{9,8,1}\oplus E_{10,4,2}.
\end{aligned}
\end{equation}
To illustrate how the correspondence between fluxes work, let us consider the IIB flux $P_{a,bc}$ that is the S-dual of Q-flux. Since these are all magnetic fluxes, we shall first perform Hodge-dualisation in 10 dimensions and write
\begin{equation}
\begin{aligned}
P_{a,bc} \longrightarrow P_{a_1\ldots a_{10},bc,a}=10\dt_{[a_1}E_{a_2\ldots a_9],bc,a}+\cdots,
\end{aligned}
\end{equation}
where the ellipses denote other contributions from gauge potentials of lower rank. Hence, we arrive at the IIB potential $E_{9,2,1}$. Similarly, we can understand the fluxes with an upper vector index, say $P^a{}_{bc}$, as 
\begin{equation}
P^a{}_{bc} \longrightarrow P_{a_1\ldots a_{9},bc}=9\dt_{[a_1}E_{a_2\ldots a_{9}],bc}+\ldots,
\end{equation}
which gives the IIB potential $E_{8,2}$. To recover the potentials of the form $E_{10,p,q}$ one must consider all the wrapping rules in their full glory and the reader is referred to \cite{Bergshoeff:2011ee}.

The wrapping rules for exotic branes tell us that all indices must take values such that one ends up with a form in the non-compact space in order to end up with a consistent compactification. In other words, if there are extra sets of indices as in $E_{9,5,1}$, the sets of indices on the right must take values in the compact dimensions and moreover be repeated. The remaining must form worldvolume directions of the corresponding brane. Hence, say for $E_{9,5,1}$ we must write
\begin{equation}
E_{a_1\ldots a_4x_1x_2x_3x_4y,x_1x_2x_3x_4y,y},
\end{equation} 

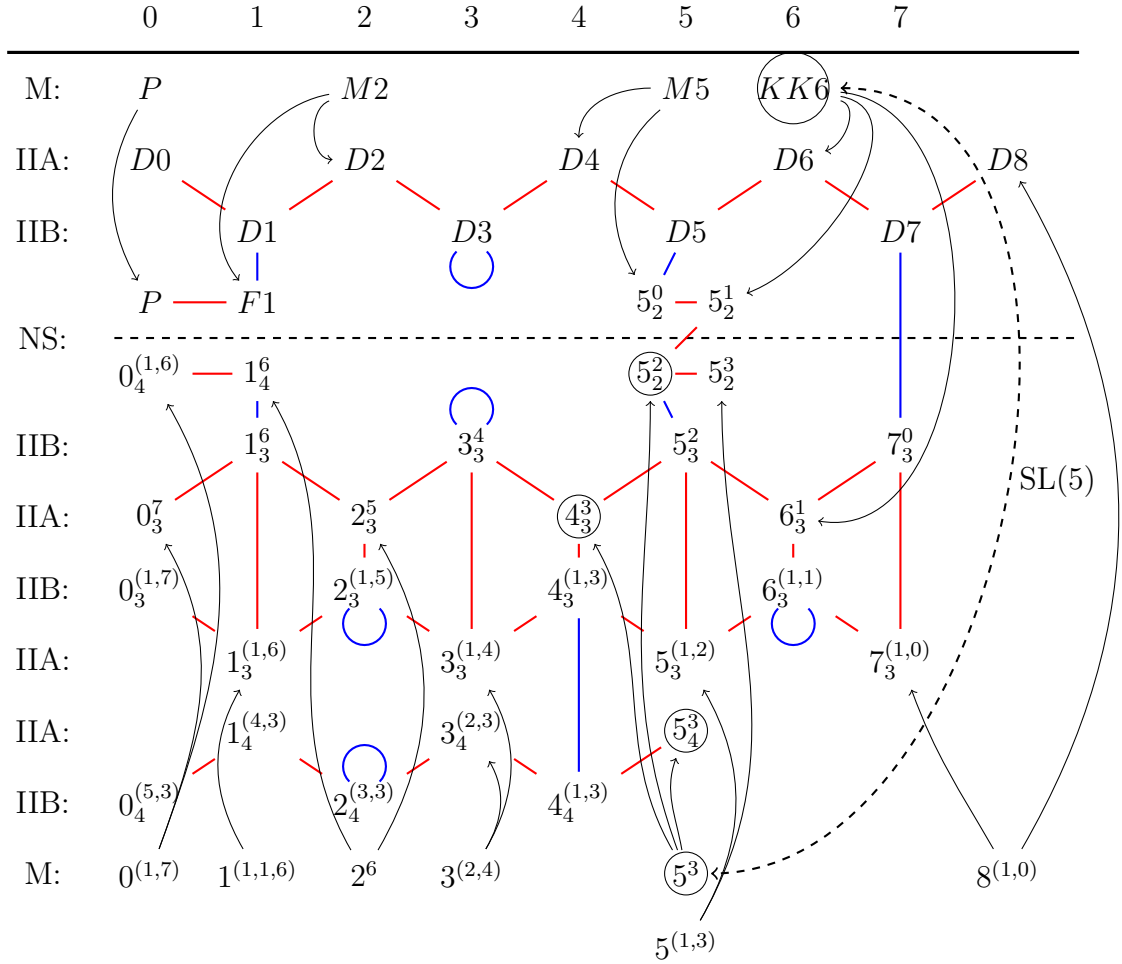
\begin{figure}[H]
	\centering
	\begin{tikzpicture}[
	scale=0.95]
	
	\foreach \y in {0,...,7} \draw (1.5*\y+2,-0.5) node{\y};
	\draw[very thick] (0,-1) -- (15,-1);
	
	\newcommand\ym{-1.5};
	\draw (0.5,\ym) node {M:}; 
	\draw (2,\ym) node (PM) {$P$}; 
	\draw (5,\ym) node (M2) {$M2$};
	\draw (9.5,\ym) node (M5) {$M5$};
	\draw (11,\ym) node (KK6) {$KK6$};
	
	\newcommand\ya{-2.5};
	\draw (0.5,\ya) node {IIA:}; 
	\draw (2,\ya) node (D0) {$D0$}; 
	\draw (5,\ya) node (D2) {$D2$};
	\draw (8,\ya) node (D4) {$D4$};
	\draw (11,\ya) node (D6) {$D6$};
	\draw (14,\ya) node (D8) {$D8$};
	
	\newcommand\yb{-3.5};
	\draw (0.5,\yb) node {IIB:}; 
	\draw (3.5,\yb) node (D1) {$D1$}; 
	\draw (6.5,\yb) node (D3) {$D3$};
	\draw (9.5,\yb) node (D5) {$D5$};
	\draw (12.5,\yb) node (D7) {$D7$};
	
	\newcommand\yns{-4.5};
	\draw (0.5,\yns-0.5) node {NS:}; 
	\draw (2,\yns) node (P) {$P$};
	\draw (3.5,\yns) node (F1) {$F1$}; 
	\draw (9,\yns) node (NS5) {$5_2^0$};
	\draw (10,\yns) node (KKM) {$5_2^1$};
	\draw[thick, dashed] (1.5,-5) -- (15,-5); 
	\draw (2,\yns-1) node (016) {$0_4^{(1,6)}$};
	\draw (3.5,\yns-1) node (146) {$1_4^6$}; 
	\draw (9,\yns-1) node (522) {$5_2^2$};
	\draw (10,\yns-1) node (523) {$5_2^3$};
	
	\newcommand\ybng{-6.5};
	\draw (0.5,\ybng) node {IIB:}; 
	\draw (3.5,\ybng) node (136) {$1_3^6$}; 
	\draw (6.5,\ybng) node (334) {$3_3^4$};
	\draw (9.5,\ybng) node (532) {$5_3^2$};
	\draw (12.5,\ybng) node (730) {$7_3^0$};
	
	\newcommand\yang{-7.5};
	\draw (0.5,\yang) node {IIA:}; 
	\draw (2,\yang) node (037) {$0_3^7$}; 
	\draw (5,\yang) node (235) {$2_3^5$};
	\draw (8,\yang) node (433) {$4_3^3$};
	\draw (11,\yang) node (631) {$6_3^1$};
	
	\newcommand\ybngt{-8.5};
	\draw (0.5,\ybngt) node {IIB:}; 
	\draw (2,\ybngt) node (0317) {$0_3^{(1,7)}$}; 
	\draw (5,\ybngt) node (2315) {$2_3^{(1,5)}$};
	\draw (8,\ybngt) node (4313) {$4_3^{(1,3)}$};
	\draw (11,\ybngt) node (6311) {$6_3^{(1,1)}$};
	
	\newcommand\yangt{-9.5};
	\draw (0.5,\yangt) node {IIA:}; 
	\draw (3.5,\yangt) node (1316) {$1_3^{(1,6)}$}; 
	\draw (6.5,\yangt) node (3314) {$3_3^{(1,4)}$};
	\draw (9.5,\yangt) node (5312) {$5_3^{(1,2)}$};
	\draw (12.5,\yangt) node (7310) {$7_3^{(1,0)}$};
	
	\newcommand\yangf{-10.5};
	\draw (0.5,\yangf) node {IIA:}; 
	\draw (3.5,\yangf) node (1443)  {$1_4^{(4,3)}$}; 
	\draw (6.5,\yangf) node (3423)  {$3_4^{(2,3)}$};
	\draw (9.5,\yangf) node (543)   {$5_4^3$};
	
	\newcommand\ybngf{-11.5};
	\draw (0.5,\ybngf) node {IIB:}; 
	\draw (2,\ybngf) node (0453) {$0_4^{(5,3)}$}; 
	\draw (5,\ybngf) node (2433) {$2_4^{(3,3)}$};
	\draw (8,\ybngf) node (4413) {$4_4^{(1,3)}$};

	\newcommand\ymng{-12.5};
	\draw (0.5,\ymng) node {M:}; 
	\draw (2,\ymng) node (017) {$0^{(1,7)}$}; 
	\draw (3.5,\ymng) node (1116) {$1^{(1,1,6)}$};
	\draw (5,\ymng) node (26) {$2^6$};
	\draw (6.5,\ymng) node (324) {$3^{(2,4)}$};
	\draw (9.5,\ymng) node (53) {$5^3$};
	\draw (9.5,\ymng-1) node (513) {$5^{(1,3)}$};
	\draw (14,\ymng) node (810) {$8^{(1,0)}$};
	
	\draw[thick, red] (D0) -- (D1) -- (D2) -- (D3) -- (D4) -- (D5) -- (D6) -- (D7) -- (D8);
	\draw[thick, red] (037) -- (136) -- (235) -- (334) -- (433) -- (532) -- (631) -- (730);
	\draw[thick, red] (NS5) -- (KKM) -- (522) -- (523) ;
	\draw[thick, red] (016) -- (146);
	\draw[thick, red] (P) -- (F1);
	\draw[thick, red] (0317) -- (1316) -- (2315) -- (3314) -- (4313) -- (5312) -- (6311) -- (7310);
	\draw[thick, red] (0453) -- (1443) -- (2433) -- (3423) -- (4413) -- (543);
	\draw[thick, red] (136) -- (1316);
	\draw[thick, red] (235) -- (2315);
	\draw[thick, red] (334) -- (3314);
	\draw[thick, red] (433) -- (4313);
	\draw[thick, red] (532) -- (5312);
	\draw[thick, red] (631) -- (6311);
	\draw[thick, red] (730) -- (7310);

	\draw[thick, blue] (D1) -- (F1);
	\draw[thick, blue] (146) -- (136); 
	\draw[thick, blue] (522) -- (532);
	\draw[thick, blue] (D5) -- (NS5);
	\draw[thick, blue] (D7) -- (730);
	\draw[thick, blue] (4313) -- (4413);
	\draw [blue,thick,domain=0:270] plot ({6.5+0.3*cos(\x+135)},{-4.0+0.3*sin(\x+135)});
	\draw [blue,thick,domain=0:270] plot ({6.5-0.3*cos(\x+135)},{-6.0-0.3*sin(\x+135)});
	\draw [blue,thick,domain=0:270] plot ({5+0.3*cos(\x+135)},{-9+0.3*sin(\x+135)});
	\draw [blue,thick,domain=0:270] plot ({5-0.3*cos(\x+135)},{-11-0.3*sin(\x+135)});
	\draw [blue,thick,domain=0:270] plot ({11+0.3*cos(\x+135)},{-9+0.3*sin(\x+135)});
	
	\path[->, black] (PM) edge[out=-120, in=120] (P);
	\path[->, black] (M2) edge[out=190, in=130] (F1);
	\path[->, black] (M2) edge[out=200, in=180] (D2);
	\path[->, black] (M5) edge[out=180, in=90] (D4);
	\path[->, black] (M5) edge[out=220, in=120] (NS5);
	\path[->, black] (KK6) edge[out=-15, in=15] (D6);
	\path[->, black] (KK6) edge[out=-10, in=20] (KKM);
	\path[->, black] (KK6) edge[out=-5, in=-10] (631);
	
	\path[->, black] (017) edge[out=70, in=-60] (016);
	\path[->, black] (017) edge[out=70, in=-60] (037);
	\path[->, black] (1116) edge[out=120, in=-120] (1316);
	\path[->, black] (26) edge[out=120, in=-60] (146);
	\path[->, black] (26) edge[out=60, in=-60] (235);
	\path[->, black] (324) edge[out=60, in=-60] (3314);
	\path[->, black] (324) edge[out=60, in=-60] (3423);
	\path[->, black] (53) edge[out= 120, in=-60] (433);
	\path[->, black] (53) edge[out= 110, in=-90] (522);
	\path[->, black] (53) edge[out= 100, in=-110] (543);
	\path[->, black] (513) edge[out=60, in=-60] (5312);
	\path[->, black] (513) edge[out=60, in=-90] (523);
	\path[->, black] (810) edge[out=120, in=-70] (7310);
	\path[->, black] (810) edge[out=60, in=-60] (D8);
	
	\path[<->, thick, black, dashed] (KK6) edge[out=0, in=0] (53);
	\draw (14.7,-7) node (U) {${\rm SL(5)}$};
	
	\draw  (522) circle [radius=0.3]; 
	\draw  (543) circle [radius=0.3]; 
	\draw  (53) circle [radius=0.3]; 
	\draw  (433) circle [radius=0.3]; 
	\draw  (KK6) circle [radius=0.5]; 
	
	\end{tikzpicture}
	\caption{\label{fig1}Branes of Type II theories whose tension is proportional to $g_s^{-\a}$ with $\a \leq 4$ and the relations between them and the M-theory branes (only some of the branes with $\a=4$ are shown). The black, red and blue arrows respectively denote reduction/oxidation, T-duality and S-duality.}
	\label{dia_br}
\end{figure}

\noindent with $x_1,\ldots,x_4$ and $y$ in the compact directions and $a_1,\ldots, a_4$ in the non-compact directions. This couples to the worldvolume of a 3-brane. To identify the brane, one notices that the counting implies that the tension of the brane coupled to $E_{b+c+d,c+d,d}$ is proportional to $d$ radii to the third power, $c$ radii to the second power and to $b$ radii entering linearly. Hence, following the conventions of \cite{deBoer:2012ma} we write
\begin{equation}
\label{conv1}
\begin{aligned}
E_{b+c+1,c} &\Longleftrightarrow b_3^{c}, \\ 
E_{b+c+d+1,c+d,d} &\Longleftrightarrow b_3^{(d,c)}
\end{aligned}
\end{equation}
with the subscript $_3$ indicating that these are $g_s^{-3}$ objects, whose masses in type II string theory are given by:
\begin{align}
\begin{aligned}
M_{b_3^{c}}&=\fr{R_{i_1}\cdots R_{i_b}R^2_{j_1}\cdots R^2_{j_c}}{g_s^3 l_s^{b+2c+1}},\\
M_{b_3^{(d,c)}}&=\fr{R_{i_1}\cdots R_{i_b}R^2_{j_1}\cdots R^2_{j_c} R^3_{k_1}\cdots R^3_{k_d}}{g_s^3 l_s^{b+2c+3d+1}} \, .
\end{aligned}
\end{align}
Starting with the T- and S-duality rules for the fluxes one can determine the corresponding rules for the branes as follows:
\begin{equation}
\label{TS}
\begin{aligned}
& T_x:&& R_x \longrightarrow \fr{l_s^2}{R_x}, && g_s \longrightarrow \fr{l_s}{R_x}g_s,\\
& S: && g_s \longrightarrow \fr{1}{g_s}, && l_s \longrightarrow g_s^{1/2}l_s.
\end{aligned}
\end{equation}
Using these rules it is straightforward to compose Figure \ref{dia_br} containing all branes with $\a \leq 3$ and some of the branes with $\a=4$.

The corresponding gauge potentials that these branes interact with and the co-dimensions $n$ of the branes have been collected in Table \ref{tab:BranesAndPotentials}.
\begin{table}[H]
\centering
\begin{tabulary}{\textwidth}{CCLLLLL}
\toprule
& & \multicolumn{2}{c}{IIA} & & \multicolumn{2}{c}{IIB}\\
\cmidrule{3-4}\cmidrule{6-7}
Co-dimension & & Potential & Brane & & Potential & Brane\\
\midrule
$n=2$ & & $E_{8,1}$ & $6_3^1$ & & $E_{8,0}$ & $7_3^0$\\
	& & $E_{8,3}$ & $4_3^3$ & & $E_{8,2}$ & $5_3^2$\\
	& & $E_{8,5}$ & $2_3^5$ & & $E_{8,4}$ & $3_3^4$\\
	& & $E_{8,7}$ & $0_3^7$ & & $E_{8,6}$ & $1_3^6$\\
$n=1$ & & $E_{9,1,1}$ & $7_3^{(1,0)}$ & & $E_{9,2,1}$ & $6_3^{(1,1)}$\\
	& & $E_{9,3,1}$ & $5_3^{(1,2)}$ & & $E_{9,4,1}$ & $4_3^{(1,3)}$\\
	& & $E_{9,5,1}$ & $3_3^{(1,4)}$ & & $E_{9,6,1}$ & $2_3^{(1,5)}$\\
	& & $E_{9,7,1}$ & $1_3^{(1,6)}$ & & $E_{9,8,1}$ & $0_3^{(1,7)}$\\
$n=0$ & & $E_{10,3,2}$ & $6_3^{(2,1)}$ & & $E_{10,2,2}$ & $7_3^{(2,0)}$\\
	& & $E_{10,5,2}$ & $4_3^{(2,3)}$ & & $E_{10,4,2}$ & $5_3^{(2,2)}$\\
	& & $E_{10,7,2}$ & $2_3^{(2,5)}$ & & $E_{10,6,2}$ & $3_3^{(2,4)}$\\
\bottomrule
\end{tabulary}
\caption{The mixed-symmetry gauge potentials that the exotic branes couple to, sorted by co-dimension $n$, for both Type IIA and Type IIB objects.}
\label{tab:BranesAndPotentials}
\end{table}


All the listed branes of Type IIA should descend from those of M-theory and so this allows one to recover the latter, given the spectrum of exotic fluxes in Type IIA. Whilst, in principle, the exotic fluxes of M-theory could be classified as components of the corresponding embedding tensor gauging the full U-duality group, there is added complexity here in that the corresponding group $E_{11}$ is not understood well enough. Hence, we shall use the following lift/reduction rules instead to conjecture new exotic branes and their corresponding gauge potentials:
\begin{equation}
\label{M2II}
\begin{aligned}
& \text{IIA} \rightarrow \text{M}: && l_s \to \fr{l_{11}^{3/2}}{R_{10}{}^{1/2}},&& g_s:\to \fr{R_{10}^{3/2}}{l_{11}^{3/2}},\\
& \text{M} \rightarrow \text{IIA}: && R_{x}\to g_s l_s,&& l_{11} \to {g_s}^{1/3} l_s,
\end{aligned}
\end{equation}
where, in the second line, the reduction is performed along a direction $x$ which is chosen to be the M-theory circle. The result will differ, depending on whether this direction is chosen to be along or transverse to the worldvolume of the brane, or along its special (quadratic or cubic) circle. Using these rules we have extended Figure \ref{dia_br} with non-geometric branes living in 11 dimensions that give rise to all branes in Type IIA with $\a\leq 3$.

\subsection{Fluxes of SL(5) EFT}
\label{exotic_intro_2}

In the previous section, it was shown how fluxes magnetically sourced by both standard and exotic branes descend from the embedding tensor of the $E_{7(7)}$ theory and how they can be organized into the tensor $\q_{MA}$ in the vector-spinor representation of $\operatorname{O}(10,10)$. The focus above was only on the branes whose tension is proportional to $g_s^{-3}$, while in general the embedding tensor of the $E_{7(7)}$ theory provides many more fluxes and generates branes with tension proportional to up to $g_s^{-7}$ \cite{Bergshoeff:2011ee,Kleinschmidt:2011vu,Lombardo:2016swq}.

It is, however, interesting to classify the branes living in M-theory whose fluxes fit a chosen U-duality group. To retain control over the proliferation of irreducible representations, let us focus on the SL(5) exceptional field theory, whose embedding tensor lives in
\begin{equation}
\Q_{mn, k}{}^l \bf \in \ol{10}\otimes 24 \rightarrow \ol{10}\oplus \ol{15}\oplus 40,
\end{equation}
where $m,n =1, \cdots, 5$ label the fundamental representation of SL(5). The 11D SUGRA is embedded in the SL(5) EFT as  $\operatorname{SL}(5) \hookleftarrow \operatorname{SL}(4)\times \RR_+$, that induces the following decomposition
\begin{equation}
\label{eq:embdec}
\begin{aligned}
\theta_{[mn]}: && \bf \ol{10} & \longrightarrow \bf 6_{-2}\oplus \ol{4}_{3}, \\
Y_{(mn)}: && \bf \ol{15} & \longrightarrow \bf \ol{10}_{-2}\oplus \ol{4}_{3}\oplus 1_{8}, \\
Z^{mn,p}: && \bf 40 & \longrightarrow \bf 4_{-7}\oplus 10_{-2}\oplus 6_{-2}\oplus 20_{3}.
\end{aligned}
\end{equation}
To represent the actual fluxes sourced by branes that fit the internal 4 dimensions of $11=4+7$ split, these representations must be combined into reducible tensorial expressions. 
We shall denote these fluxes as in the following and we also indicate the some of the branes that contribute to these fluxes. The exotic branes that we shall be mainly concerned with can contribute to several of these components as we will see in an explicit example. Our notation for the fluxes is as follows
\begin{equation}
\label{eq:fluxbranes}
\begin{aligned}
\bf 1_8:&\ F_{abcd}, && \text{M5}\\
\bf 20_{3}\oplus \ol{4}_3^{\it (Y)}: &\ Z^{ab,c}=\e^{abef}f_{ef}{}^{c}-5\e^{efc[a}f_{ef}{}^{b]}, && \text{KK6}\\
&\ T_a=f_{ab}{}^{b},\\
\bf \ol{10}_{-2}+6_{-2}^{\it (Z)}:&\ Q_{(a,b)}+Q_{[a,b]}=Q_{a,b}=\e_{bcde}Q_a{}^{cde} && 
\\
\bf \bf 4_{-7}:&\ L^a &&
\\
\bf 10_{-2}+6_{-2}^{\it(\q)}:&\ \mT^{(a,b)}+\mT^{[a,b]}=\mT^{a,b} && 
\end{aligned}
\end{equation}
where we employ the notation of \cite{Blair:2014zba} that the letter in superscript of an irreducible representation denotes its origin and where confusion may occur and indices $a=1,\cdots,4$ label the directions of the M-theory torus. The remaining $\bf \ol{4}_3^{\it \theta}$ coming from the trombone gauging does not correspond to a BPS brane.

In principle these fluxes may then be expressed in terms of potential fields, say $F_{abcd}=4\dt_{[a}C_{bcd]}+\ldots$ where the ellipses denote additional terms which deform the na\"{i}ve first term into the full flux sitting in a U-duality irreducible representation. The precise form of the expressions depend on the frame chosen, which is a U-duality analogue of the switch between the B-field and the $\b$-field in DFT. Hence, to escape such ambiguities, we shall always associate the irreducible representations with the fluxes rather than the potentials.

\section{Orbits of the SL(5) U-duality group}
\label{orbits}

The components of the generalised torsion of SL(5) EFT are irreducible representations that can be characterized by their weight diagrams. The Lie algebra $\mathfrak{sl}(5)$ maps from the different weight spaces to one another. A subset of the $\mathfrak{sl}(5)$ transformations are the usual infinitesimal diffeomorphisms and they do not rotate different types of fluxes as in the decomposition~\eqref{eq:embdec} into each other. 

We use the following strategy to determine how the U-duality group acts on fluxes. First, one determines the simple root of a U-duality group that becomes decoupled under the embedding of $G \hookleftarrow \operatorname{GL}(d)$, where $G$ is the corresponding U-duality group and $\operatorname{GL}(d)$ represents internal diffeomorphisms.  Next, one constructs weight diagrams for each of the irreducible representations contributing to the embedding tensor and then determines the branching into irreducible representation of $\operatorname{GL}(d)$. Finally, the connections between the weights of different irreducible representations of $\operatorname{GL}(d)$ inside a given irreducible representation of $G$, generated by the decoupled root, will precisely give the U-duality orbits. We shall demonstrate this procedure by applying this idea to the SL(5) EFT.

\subsection{Root system of SL(5)}

The simple roots of the Lie algebra $\mathfrak{sl}(5)$ in the canonical $\w$-basis of fundamental weights are given by the following
\begin{equation}
\begin{aligned}
\a_{12}&=(2,-1,0,0),\\
\a_{23}&=(-1,2,-1,0),\\
\a_{34}&=(0,-1,2,-1),\\
\a_{45}&=(0,0,-1,2),
\end{aligned}
\end{equation}
where the labelling of the roots will become clear momentarily. The remaining positive roots are
\begin{equation}
\begin{aligned}
&\a_{13}=\a_{12}+\a_{23}, \qquad \a_{14}=\a_{12}+\a_{23}+\a_{34}, \\
&\a_{24}=\a_{23}+\a_{34}, \qquad \a_{25}=\a_{23}+\a_{34}+\a_{45},\\
&\a_{35}=\a_{34}+\a_{45}, \qquad \a_{15}=\a_{12}+\a_{23}+\a_{34}+\a_{45}.
\end{aligned}
\end{equation}
In addition, one has the same  number of negative roots and four Cartan generators. We shall use the simple root $\alpha_{45}$ to be the one associated with the decomposition $\operatorname{SL}(5)\to \operatorname{GL}(4)$.

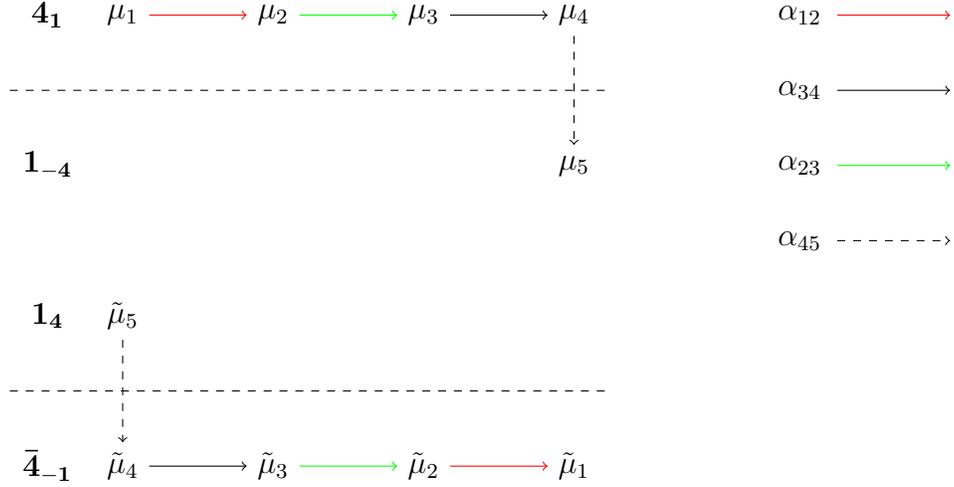
\begin{figure}[ht]
\centering
\begin{tikzpicture}
 

\draw (1,0) node (1) {$\m_1$}; 
\draw (3,0) node (2) {$\m_2$};
\draw (5,0) node (3) {$\m_3$};
\draw (7,0) node (4) {$\m_4$}; 
\draw (7,-2) node (5) {$\m_5$};

\draw (0,-2) node {$\bf 1_{-4}$} ;
\draw (0,0) node {$\bf {4}_{1}$} ;


\draw (1,-4) node (5a) {$\tilde\m_5$}; 
\draw (1,-6) node (4a) {$\tilde\m_4$}; 
\draw (3,-6) node (3a) {$\tilde\m_3$};
\draw (5,-6) node (2a) {$\tilde\m_2$};
\draw (7,-6) node (1a) {$\tilde\m_1$};

\draw (0,-4) node {$\bf 1_{4}$} ;
\draw (0,-6) node {$\bf {\bar{4}}_{-1}$} ;

\draw[dashed] (-0.5,-1) -- (7.5,-1);

\draw[->, dashed] (4) edge (5);
\draw[->] (3) edge (4);
\draw[->, red] (1) edge (2);
\draw[->, green] (2) edge (3) ;

\draw[dashed] (-0.5,-5) -- (7.5,-5);

\draw[->, dashed] (5a) edge (4a);
\draw[->] (4a) edge (3a);
\draw[->, red] (2a) edge (1a);
\draw[->, green] (3a) edge (2a) ;

\newcommand\zr{10}
\draw (\zr,0) node  {$\a_{12}$};
\draw (\zr,-1) node {$\a_{34}$};
\draw (\zr,-2) node {$\a_{23}$};
\draw (\zr,-3) node {$\a_{45}$};
\draw[->, red] (\zr+0.5,0) -- (\zr+2,0);
\draw[->] (\zr+0.5,-1) -- (\zr+2,-1);
\draw[->, green] (\zr+0.5,-2) -- (\zr+2,-2);
\draw[->, dashed] (\zr+0.5,-3) -- (\zr+2,-3);

\end{tikzpicture}
\caption{Weight diagram of the fundamental  $\bf 5$ and the anti-fundamental $\bf{\bar{5}}$ representations of $\mathfrak{sl}(5)$ with weights arranged into irreducible representations under the decomposition $\operatorname{SL}(5) \to \operatorname{SL}(4) \times \operatorname{GL}(1)$ when the root $\a_{45}$ is deleted. The highest weights here are $\m_{1}$ and $\tilde\m_5$ respectively. The action of different roots is denoted by different colours and the direction of arrows shows the lowering of the weight.}
\label{5}
\end{figure}

A convenient  basis to produce the appropriate SL(5) rotation matrices for generating the non-geometric backgrounds later is as follows
\begin{equation}
\begin{aligned}
\m_1=
\begin{bmatrix}
1 \\ 0 \\0 \\0 \\0 
\end{bmatrix}, && 
\m_2=\begin{bmatrix}
0 \\ 1 \\0 \\0 \\0 
\end{bmatrix}, &&
\m_3=\begin{bmatrix}
0 \\ 0 \\1 \\0 \\0 
\end{bmatrix}, &&
\m_4=\begin{bmatrix}
0 \\ 0 \\0 \\1 \\0 
\end{bmatrix}, &&
\m_5=\begin{bmatrix}
0 \\ 0 \\0 \\0 \\1 
\end{bmatrix},
\end{aligned}
\end{equation}
where $\m_m$ are the basis vectors of the fundamental representation which all lie in different weight spaces. A vector in this representation is expanded as $V=V^m\m_m$. In this basis the action of a root $-\a_{mn}$ (for some fixed $m$ and $n$) will transform the vector component $V^m$ into $V^n$. A similar notation will be employed for the conjugate co-vectors. Since the Lie algebra elements act as a derivative, the action of the root $-\a_{mn}$ on a tensor will produce a sum of terms with the corresponding index conversion in each term. For example one can write
\begin{equation}
\begin{aligned}
&E_{-\a_{35}}:&& V^{3} \longrightarrow V^5, \\
& && V_{5} \longrightarrow -V_3,\\
& && T^{53} \longrightarrow T^{55} ,\\
& && T^{33} \longrightarrow T^{53}+T^{35},
\end{aligned}
\end{equation}
where one notices that the dual representation transforms in the opposite way as it should. The signs on the right-hand sides will be of little relevance for our discussion.

Using the weight diagram Figure \ref{5}, it is straightforward to recover the explicit form of the matrix corresponding to SL(2) rotations generated by the subalgebra $(\a_{45}, -\a_{45})$ and the corresponding Cartan generator. Explicitly, we have
\begin{equation}
U_{123}[\f]=\begin{bmatrix}
1 & 0 & 0 & 0 & 0\\
0 & 1 & 0 & 0 & 0\\
0 & 0 & 1 & 0 & 0\\
0 & 0 & 0 & \cos \f & \sin \f \\
0 & 0 & 0 & -\sin \f & \cos \f\\
\end{bmatrix} \in \operatorname{SL}(5),
\end{equation}
which rotates the vectors $\m_4$ and $\m_5$ into each other. The notation $U_{123}$ was chosen to comply with that of \cite{Blair:2014zba}. In their notation, one may understand the subscript numbers as corresponding to the weights orthogonal to the rotation plane. To show that the above matrix indeed represents a rotation in the plane $(\m_5,\m_4)$ we first notice that these weights transform in the fundamental representation of $\operatorname{SL}(2)$ under the action of the roots $\a_{45}, -\a_{45}$ and the corresponding Cartan generator. Hence, one writes 
\begin{equation}
\begin{aligned}
E_{\a_{45}}=
\begin{bmatrix}
0 & 1 \\
0 & 0
\end{bmatrix}, \qquad
E_{-\a_{45}}=
\begin{bmatrix}
0 & 0 \\
1 & 0
\end{bmatrix},
\end{aligned}
\end{equation}
and the lower right block of the matrix $U_{123}$ is simply obtained as
\begin{equation}
\label{u123}
u_{123}[\f]=\exp\Big[\f (E_{\a_{45}}-E_{-\a_{45}})\Big].
\end{equation}
We shall see that the transformation that transforms KK6 into an exotic brane is given by $u_{123}[\p/2]$ and reads
\begin{equation}
u_{123}=\begin{bmatrix}
0 & 1 \\
-1 & 0
\end{bmatrix}.
\end{equation}
Hence, this group element simply replaces $V^4 \leftrightarrow -V^5$ and more generally it interchanges all indices $4\leftrightarrow 5$ on any SL(5) tensor (up to sign). As is common for elements of the form \eqref{u123} with $\varphi=\frac{\pi}{2}$ the finite duality transformation is (the covering of) a Weyl transformation, namely an elementary transposition in the symmetric group $S_5$.

\subsection{Orbits inside the \texorpdfstring{$\bf \ol{15}$}{15} representation}

Let us start with the representation $\bf \ol{15}$ that contains the fluxes $F_{abcd}$, $Q_{(a,b)}$ and the geometric flux $f_{ab}{}^{a}$. Note, that although it is tempting to identify these fluxes with particular branes, this is not a one-to-one correspondence. Indeed, already for the 5-brane orbit of DFT one notices, that the Q-monopole sources not only the flux $Q_a{}^{bc}$ but also the geometric flux $f_{ab}{}^a$ \cite{Bakhmatov:2016kfn}. The same is true for the H-monpole.  This is related to the fact that we consider the full 11D solutions rather than their truncated 7D or 4D versions. 

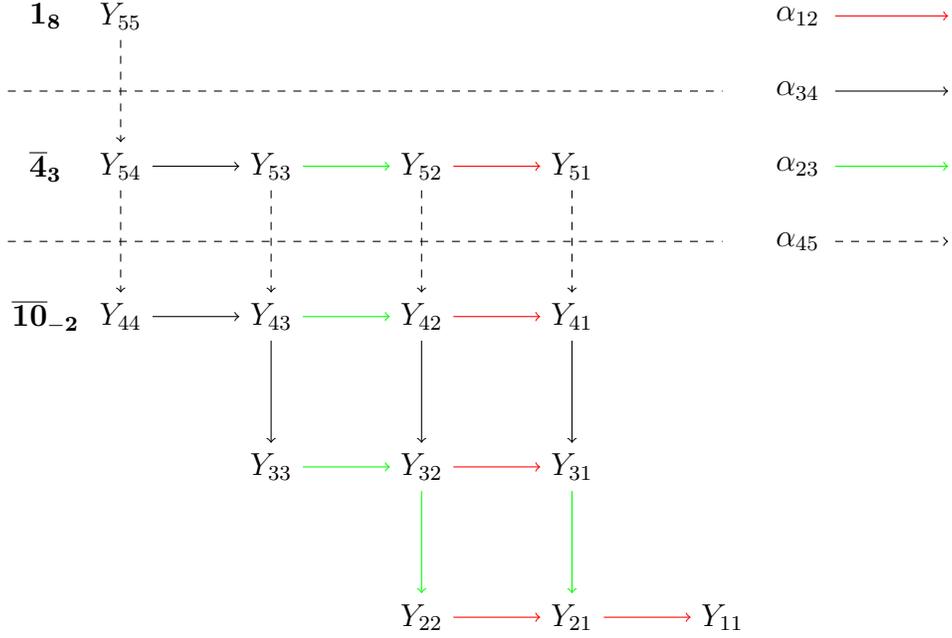
\begin{figure}[ht]
\centering
\begin{tikzpicture}

\draw (1,0) node (1)  {$Y_{55}$}; 
\draw (1,-2) node (2) {$Y_{54}$}; 
\draw (3,-2) node (4) {$Y_{53}$};
\draw (5,-2) node (6) {$Y_{52}$};
\draw (7,-2) node (7) {$Y_{51}$}; 
\draw (1,-4) node (3) {$Y_{44}$}; 
\draw (3,-4) node (5) {$Y_{43}$}; 
\draw (5,-4) node (9) {$Y_{42}$}; 
\draw (7,-4) node (10){$Y_{41}$}; 
\draw (3,-6) node (8) {$Y_{33}$};
\draw (5,-6) node (11){$Y_{32}$}; 
\draw (7,-6) node (12){$Y_{31}$};
\draw (5,-8) node (13){$Y_{22}$}; 
\draw (7,-8) node (14){$Y_{21}$}; 
\draw (9,-8) node (15){$Y_{11}$};

\draw (0,0) node {$\bf 1_8$} ;
\draw (0,-2) node {$\bf \ol{4}_{3}$} ;
\draw (0,-4) node {$\bf \ol{10}_{-2}$} ;

\draw[->, dashed] (1) edge (2) (2) edge (3) (4) edge (5) (6) edge (9) (7) edge (10);
\draw[->] (2) edge (4) (3) edge (5) (5) edge (8)  
(9) edge (11) (10) edge (12);
\draw[->, red] (6) edge (7) (9) edge (10) (11) edge (12) (14) edge (15) (13) edge (14);
\draw[->, green] (4) edge (6) (5) edge (9) (8) edge (11) (11) edge (13) (12) edge (14);

\draw[dashed] (-0.5,-1) -- (9,-1);
\draw[dashed] (-0.5,-3) -- (9,-3);

\newcommand\zr{10}
\draw (\zr,0) node  {$\a_{12}$};
\draw (\zr,-1) node {$\a_{34}$};
\draw (\zr,-2) node {$\a_{23}$};
\draw (\zr,-3) node {$\a_{45}$};
\draw[->, red] (\zr+0.5,0) -- (\zr+2,0);
\draw[->] (\zr+0.5,-1) -- (\zr+2,-1);
\draw[->, green] (\zr+0.5,-2) -- (\zr+2,-2);
\draw[->, dashed] (\zr+0.5,-3) -- (\zr+2,-3);

\end{tikzpicture}
\caption{Weight diagram of the $\bf \ol{15}$ of $\mathfrak{sl}(5)$ with weights arranged into irreducible representations upon decomposition $\operatorname{SL}(5) \to \operatorname{SL}(4) \times \operatorname{GL}(1)$ when the root $\a_{45}$ is deleted. The action of different roots is denoted by different colours and the direction of an arrow shows the lowering of a weight.}
\label{15}
\end{figure}

The structure of this representation is schematically depicted in Figure \ref{15}.

Since the branching $\operatorname{SL}(5) \to \operatorname{SL}(4) \times \operatorname{GL}(1)$ is obtained upon deleting the root $\a_{45}$, it is not surprising that this root allows one to travel between different irreducible representations of $\operatorname{SL}(4)$ inside SL(5). In other words, the corresponding operator $U_{123}$ is precisely the desired U-duality transformation that transforms the flux $F_{abcd}$ into the Q-flux and the trace part of the geometric flux $f_{ab}{}^c$.

To understand the precise action of the transformation $U_{123}$ on the weights of Figure \ref{15}, one first notices that the components $(Y_{55},Y_{54},Y_{44})$ transform as a three-dimensional representation of the $\operatorname{SL}(2)$ subgroup generated by $\a_{45}$ and $-\a_{45}$ and the Cartan element, while the three pairs $(Y_{i5},Y_{i4})$ with $i=1,2,3$ all transform in the fundamental representation. Hence, for the latter, one simply concludes that the transformation just rotates the weight in each pairs into each other. Hence, we have the following U-duality orbit for each of the pairs in the fundamental representation
\begin{equation}
U_{123}: Y_{i5} \longrightarrow Y_{i4}.
\end{equation}
The three-dimensional representation is more subtle and, to derive the orbit, one first writes the operators $E_{\a_{45}}$ and $E_{-\a_{45}}$ as
\begin{equation}
\begin{aligned}
E_{\a_{45}}=
\begin{bmatrix}
0 & \sqrt{2} & 0 \\
0 & 0 & \sqrt{2} \\
0 & 0 & 0
\end{bmatrix}, && 
E_{-\a_{45}}=
\begin{bmatrix}
0 & 0 & 0 \\
\sqrt{2} & 0 &0 \\
0 & \sqrt{2} & 0
\end{bmatrix}.
\end{aligned}
\end{equation}
Now, simply calculating $u_{123}$ and setting $\f=\p/2$, we obtain the following matrix
\begin{equation}
\label{u3}
u_{123}=
\begin{bmatrix}
0 & 0 & 1 \\
0& -1 & 0 \\
1& 0 & 0
\end{bmatrix}.
\end{equation}
This implies, that the U-duality transformation $U_{123}$ for $\f=\p/2$ rotates $Y_{55}$ into $Y_{44}$, while $Y_{54}$ just acquires a prefactor. This is in agreement with the general property that $U_{123}$ is a Weyl transformation exchanging the indices $4$ and $5$ (up to sign). In terms of flux components one writes
\begin{equation}
\begin{aligned}
& U_{123}: && Y_{55}=F_{1234} && \longrightarrow Y_{44}=Q_{(4,4)}, \\
& && Y_{54}=f_{a4}{}^{a} && \longrightarrow f_{a4}{}^{a}. \\
& && Y_{5i}=f_{ai}{}^{a} && \longrightarrow Y_{4i}=Q_{(4,i)}.
\end{aligned}
\end{equation}
These U-duality orbits are indeed consistent with the ones obtained in \cite{Blair:2014zba} for a co-dimension two brane. 



\subsection{Orbits inside the \texorpdfstring{$\bf 40$}{40} representation}

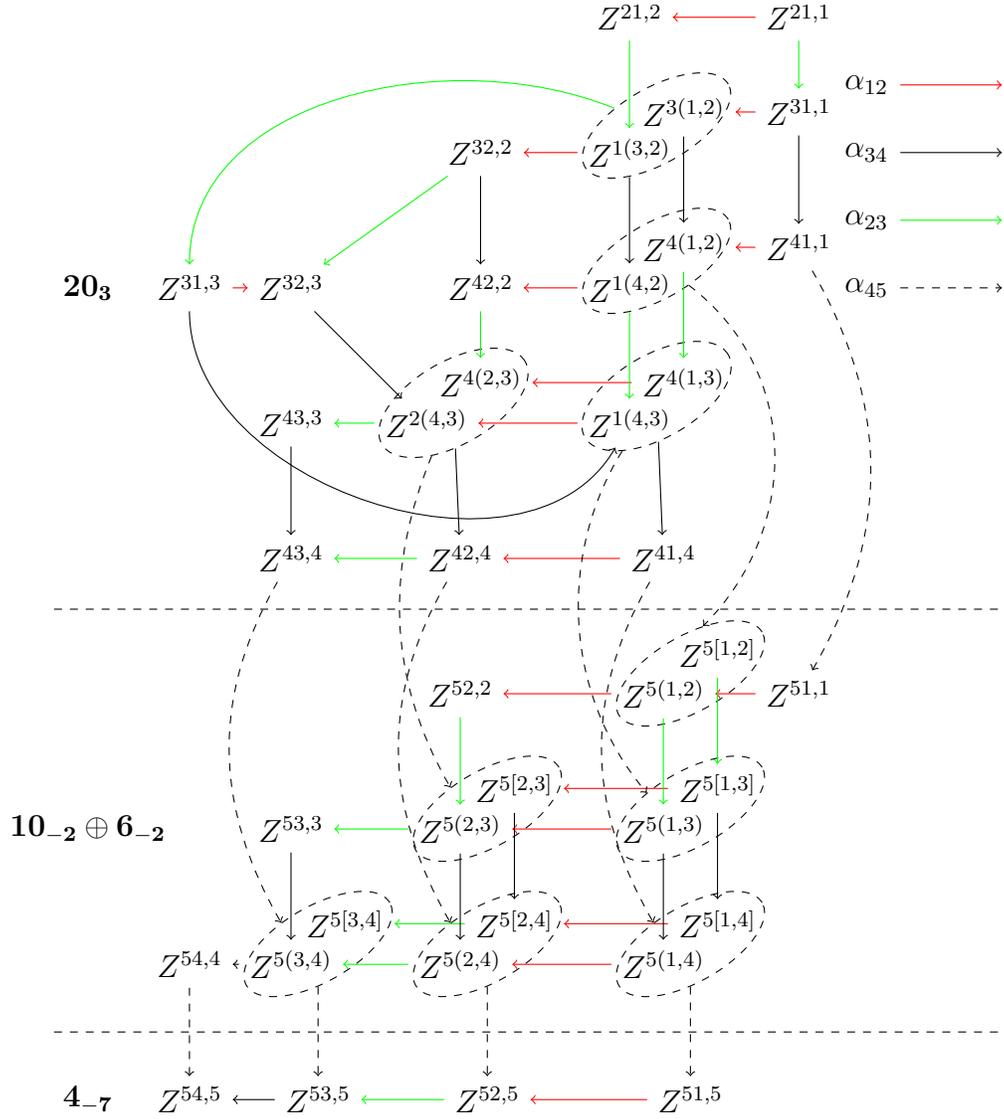
\begin{figure}[H]
\centering
\begin{tikzpicture}[scale=0.9]

\newcommand\sx{0.8}
\newcommand\sy{0.6} 

\draw (1.5,-4) node {$\bf 20_{3}$};
\draw (1.5,-12) node {$\bf {10}_{-2} \oplus 6_{-2}$};
\draw (1.5,-16) node {$\bf {4}_{-7}$};

\draw (12,0) node (1) {$Z^{21,1}$}; 
\draw (9.5,0) node (2) {$Z^{21,2}$}; 
\draw (12,-2+\sy) node (3) {$Z^{31,1}$};
\draw (9.5,-2) node (5) {$Z^{1(3,2)}$};
\draw (9.5+\sx/2,-2+\sx/2) node[ellipse, dashed, minimum height=1cm,minimum width=2.2cm,rotate=30, draw] (45) {};
\draw (9.5,-4) node (9) {$Z^{1(4,2)}$};
\draw (9.5,-6) node (17) {$Z^{1(4,3)}$};
\draw (9.5+\sx,-2+\sy) node (4) {$Z^{3(1,2)}$};
\draw (9.5+\sx,-4+\sy) node (8) {$Z^{4(1,2)}$};
\draw (9.5+\sx/2,-4+\sx/2) node[ellipse, dashed, minimum height=1cm,minimum width=2.2cm,rotate=30, draw] (89) {};
\draw (9.5+\sx,-6+\sy) node (16) {$Z^{4(1,3)}$};
\draw  (9.5+\sx/2,-6+\sx/2) node[ellipse, dashed, minimum height=1.1cm,minimum width=2.2cm,rotate=30,draw] (1617) {};
\draw (10,-8) node (23) {$Z^{41,4}$};
\draw (6.5+\sx,-2) node (7) {$Z^{32,2}$};
\draw (6.5+\sx,-4) node (12) {$Z^{42,2}$};
\draw (6.5,-6) node (20) {$Z^{2(4,3)}$};
\draw (6.5+\sx,-6+\sy) node (19) {$Z^{4(2,3)}$};
\draw (6.5+\sx/2,-6+\sy/2) node[ellipse, dashed, minimum height=1.1cm,minimum width=2.2cm,rotate=30,draw] (1920) {};
\draw (7,-8) node (26) {$Z^{42,4}$};
\draw (12,-4+\sy) node (6) {$Z^{41,1}$};
\draw (4.5,-6) node (27) {$Z^{43,3}$};
\draw (4.5,-8) node (33) {$Z^{43,4}$};
\draw (3,-4) node (10) {$Z^{31,3}$};
\draw (4.5,-4) node (13) {$Z^{32,3}$};

\draw (12,-10) node (11) {$Z^{51,1}$}; 
\draw (10,-10) node (14) {$Z^{5(1,2)}$};
\draw (7,-10) node (18) {$Z^{52,2}$};  
\draw (10,-12) node (21) {$Z^{5(1,3)}$}; 
\draw (7,-12) node (24) {$Z^{5(2,3)}$}; 
\draw (4.5,-12) node (32) {$Z^{53,3}$}; 
\draw (10,-14) node (28) {$Z^{5(1,4)}$}; 
\draw (7,-14) node (30) {$Z^{5(2,4)}$}; 
\draw (4.5,-14) node (36) {$Z^{5(3,4)}$}; 
\draw (3,-14) node (39) {$Z^{54,4}$};

\draw (10+\sx,-10+\sy) node (15) {$Z^{5[1,2]}$};
\draw (10+\sx,-12+\sy) node (22) {$Z^{5[1,3]}$}; 
\draw (7+\sx,-12+\sy) node (25) {$Z^{5[2,3]}$}; 
\draw (10+\sx,-14+\sy) node (29) {$Z^{5[1,4]}$}; 
\draw (7+\sx,-14+\sy) node (31) {$Z^{5[2,4]}$}; 
\draw (4.5+\sx,-14+\sy) node (37) {$Z^{5[3,4]}$}; 

\draw (10+\sx/2,-10+\sy/2) node[ellipse, dashed, minimum height=1cm,minimum width=2.2cm,rotate=30, draw] (1415) {};
\draw (10+\sx/2,-12+\sy/2) node[ellipse, dashed, minimum height=1cm,minimum width=2.2cm,rotate=30, draw] (2122) {};
\draw (7+\sx/2,-12+\sy/2) node[ellipse, dashed, minimum height=1cm,minimum width=2.2cm,rotate=30, draw] (2425) {};
\draw (10+\sx/2,-14+\sy/2) node[ellipse, dashed, minimum height=1cm,minimum width=2.2cm,rotate=30, draw] (2829) {};
\draw (7+\sx/2,-14+\sy/2)  node[ellipse, dashed, minimum height=1cm,minimum width=2.2cm,rotate=30, draw] (3031) {};
\draw (4.5+\sx/2,-14+\sy/2) node[ellipse, dashed, minimum height=1cm,minimum width=2.2cm,rotate=30, draw] (3637) {};

\draw (10+\sx/2,-16) node (34) {$Z^{51,5}$}; 
\draw (7+\sx/2,-16) node (35) {$Z^{52,5}$}; 
\draw (4.5+\sx/2,-16) node (38) {$Z^{53,5}$}; 
\draw (3,-16) node (40) {$Z^{54,5}$};

\draw[->, green] (45) edge[out=150, in=90] (10);
\draw[->] (10) edge[out=-90, in=-120] (17);

\draw[->, red] 
(1) edge (2) (3) edge (4) (5) edge (7) (6) edge (8) (23) edge (26)    (9) edge (12)   (10) edge (13) (11) edge (14) (14) edge (18) (16) edge (19) (17) edge (20)
       (21) edge (24) (22) edge (25) (28) edge (30)  (29) edge (31)  (34)  edge (35) ;

\draw[->, green] 
 (2) edge (5)
(12) edge (19) (20) edge (27)
(1) edge (3)  (8) edge (16) (9) edge (17)
   (26) edge (33)
 (7) edge (13) (18) edge (24) (24) edge (32) (14) edge (21) (15) edge (22) (30) edge (36) (31) edge (37) (35)  edge (38);

\draw[->] 
(7) edge (12) 
 (27) edge (33)
 (3) edge (6) 
 (1617) edge (23)   (5) edge (9)
  (4) edge (8)  (1920) edge (26) (13) edge (20)  (32) edge (36) (36) edge (39)  (21) edge (28) (22) edge (29)  (24) edge (30)  (25) edge (31) (38)  edge (40);

\draw[->, dashed] (39) edge (40) (3637) edge (38) (3031) edge (35) (2829) edge (34);

\draw[->, dashed] (89) edge[out=-45, in=45] (1415);
\draw[->, dashed] (6) edge[out=-60, in=60] (11);
\draw[->, dashed] (1617) edge[out=-130, in=130] (2122);
\draw[->, dashed] (1920) edge[out=-120, in=120] (2425);

\draw[->, dashed] (33) edge[out=-120, in=120] (3637);
\draw[->, dashed] (26) edge[out=-120, in=120] (3031);
\draw[->, dashed] (23) edge[out=-120, in=120] (2829);

\draw[dashed] (1,-8.75) -- (15,-8.75);
\draw[dashed] (1,-15) -- (15,-15);

\newcommand\zr{13}
\draw (\zr,-1) node  {$\a_{12}$};
\draw (\zr,-2) node {$\a_{34}$};
\draw (\zr,-3) node {$\a_{23}$};
\draw (\zr,-4) node {$\a_{45}$};
\draw[->, red] (\zr+0.5,-1) -- (\zr+2,-1);
\draw[->] (\zr+0.5,-2) -- (\zr+2,-2);
\draw[->, green] (\zr+0.5,-3) -- (\zr+2,-3);
\draw[->, dashed] (\zr+0.5,-4) -- (\zr+2,-4);

\end{tikzpicture}
\caption{Weight diagram of the $\bf 40$ of $\mathfrak{sl}(5)$ with weights arranged into irreducible representations upon the decomposition $\operatorname{SL}(5) \to \operatorname{SL}(4) \times \operatorname{GL}(1)$ when the root $\a_{45}$ is deleted. The components $Z^{ab,c}$ with non-zero symmetric and antisymmetric parts in $\{b,c\}$ correspond to vectors of the same weight which are collected in dashed ellipses. In the case of $\bf 10_{-2} \oplus 6_{-2}$ the symmetric and antisymmetric parts belong to different representations of $\operatorname{SL}(4)$.  The action of different roots is again denoted by different colours and the direction of the arrows show the lowering of a weight.}
\label{40}
\end{figure}

The $\bf \ol{15}$ component of the embedding tensor $\Q_{mn,k}{}^l$ contains fluxes sourced by the M5,  the KK-monopole and exotic branes. In order to get the complete spectrum of all branes, one must consider the other embedding tensor representations as well. Hence, we now turn to the representation ${\bf 40}$ of SL(5) whose weight diagram is depicted on the Figure \ref{40}.

The irreducible representation $\bf 40$ of SL(5) has the subtlety that 10 pairs of vectors of this space have the same weights i.e.\ there are non-trivial weight multiplicities. This is related to the fact that the components $Z^{ab,c}$ can be symmetric or antisymmetric in $\{b,c\}$ but still possess the same weight. Hence, depending on the index arrangement, these pairs either stay entirely in one irreducible representation or break into two vectors living in different irreducible representations under branching.

 For example, the components $Z^{4(1,2)}$ and $Z^{4[1,2]}$ both belong to the $\bf 20_3$, while the components $Z^{5(1,2)}$ and $Z^{5[1,2]}$ descend into the $\bf 10_{-2}$ and $\bf 6_{-2}$ respectively. Using the fact that each root of the algebra acts as a derivative, it is straightforward to construct the diagram in Figure \ref{40}, which explicitly shows the action of the simple roots on each weight\footnote{Strictly speaking the arrows on all the weight diagrams here correspond to actions of the negative counterparts of the simple roots, i.e.\ of $-\a$ for each simple root $\a$.}. There is still a subtlety relating to the action of the positive roots corresponding to raising operators. The action of $-\a_{23}$ on $Z^{21,2}$ gives $Z^{31,2}+Z^{21,3}=-2Z^{1(3,2)}$, up to some prefactors. Hence, according to Figure \ref{40}, one cannot lower the weight $Z^{21,2}$ to $Z^{3(2,1)}$ by the action of $-\a_{23}$. However, a straightforward check shows that raising both the weights  $Z^{3(2,1)}$ and $Z^{1(3,2)}$ gives just $Z^{21,2}$. Hence, it is necessary to keep this in mind and always keep track which plane is used for a U-duality rotation generated by a pair of roots $(\a,-\a)$.

Taking this subtlety into account, we write the following for the relevant root $\a_{45}$
\begin{equation}
\begin{aligned}
 Z^{54,5} & \longleftrightarrow  Z^{54,4}, \\
 Z^{4(i,j)} & \longleftrightarrow Z^{5(i,j)},\\
 Z^{4[i,j]} & \longleftrightarrow Z^{5[i,j]}, \\
 Z^{4i,4} & \longleftrightarrow \big(Z^{5(i,4)}+3Z^{5[i,4]}\big) \longleftrightarrow Z^{5i,5}, \\
Z^{54,i} =\big(Z^{5(i,4)}-Z^{5[i,4]}\big) & \longleftrightarrow  \big(Z^{5(i,4)}-Z^{5[i,4]}\big) \quad \textrm{(anti-invariant)},\\
Z^{5i,4} &\longleftrightarrow Z^{4i,5}=Z^{5i,4}+Z^{45,i},\\
Z^{ij,k} &\longleftrightarrow Z^{ij,k}.
\end{aligned}
\end{equation}
Similar to the $\bf \ol{15}$ representation, we observe both doublets and triplets here with respect to the action of the $\operatorname{SL}(2)$ generated by the $\a_{45}$ root. In addition, one has a scalar $\big(Z^{5(i,4)}-Z^{5[i,4]}\big)$ which corresponds to a self-dual combination of the $\mT$ and $Q$ flux components.

Combining these rules with the rules from the previous subsection in~\eqref{eq:fluxbranes} one can write the full U-duality orbit under the action of the operator $U_{123}$ for $\f=\p/2$ as
\begin{equation}
\begin{aligned}
F_{1234} &\longleftrightarrow Q^{4,4}, & Z^{4(i,j)} & \longleftrightarrow \mT^{(i,j)}, & Q_{(i,j)}& \longleftrightarrow Q_{(i,j)},\\
f_{a4}{}^a &\longleftrightarrow  f_{a4}{}^a, & \mT^{(4,4)}  &\longleftrightarrow L^{4}, & Z^{ij,k} & \longleftrightarrow Z^{ij,k}.\\
f_{ai}{}^a & \longleftrightarrow Q_{(4,i)}, & Z^{4i,4} & \longleftrightarrow L^{i}, &\mT^{(i,j)} &\longleftrightarrow Z^{4(i,j)}\\
 Z^{4[i,j]} & \longleftrightarrow \epsilon^{ijk} Q_{[4,k]}, & \mT^{(i,4)} & \longleftrightarrow \mT^{(i,4)}+\epsilon^{ijk}Q_{[j,k]}.
\end{aligned}
\end{equation}
The component $Q_{(i,j)}$ transforms to itself under this U-duality. 
Note that the components of the traceless part of the geometric $f$-flux are encoded inside $Z^{ab,c}$. Taking this into account and redefining $\{x=x^1,y=x^2, w=x^3, z=x^4\}$ one finds these U-duality orbits are in perfect agreement with what has been found in \cite{Blair:2014zba}. 


In the following section, we will focus on a field realization of the above U-duality orbits and construct the corresponding solutions in EFT. In particular we consider a KK6 monopole that fills the external space, compute its fluxes and compute a U-duality transform of it.

\section{Uplift of SUGRA solutions into the \texorpdfstring{$\operatorname{SL}(5)$}{SL(5)} EFT}
\label{SL5EFT}

We now consider explicit solutions of the SL(5) EFT and their duality orbits.

\subsection{\texorpdfstring{$\operatorname{SL}(5)$}{SL(5)} theory}

Let us start with the explicit form of the generators and projectors for the fundamental and two-form representation:
\begin{equation}
\begin{aligned}
(t^{m}{}_{p})^{k}{}_{l} &=\delta^{m}{}_{l}\delta^{k}{}_{p}-\frac{1}{5}\delta^{m}{}_{p}\delta^{k}{}_{l},\\
(T^{m}{}_{n})^{pq}{}_{rs}& =2 (t^{m}{}_{n})^{[p}{}_{[r}\delta^{q]}{}_{s]},\\
\mathbb{P}^k{}_l{}^{m}{}_{n}&= (t^r{}_s)^m{}_n(t^{s}{}_{r})^k{}_l,\\
\mathbb{P}^{mn}{}_{pq}{}^{rs}{}_{kl}&= \frac{1}{3}(T^u{}_v)^{mn}{}_{pq}(T^{v}{}_{u})^{rs}{}_{kl},\\
\mathbb{P}^k{}_l{}^{mn}{}_{pq}&=\frac{1}{\sqrt{3}} (t^r{}_s)^k{}_l(T^s{}_r)^{mn}{}_{pq}.
\end{aligned}
\end{equation}
where we have used $t^m{}_n$ to denote the generators in the fundamental and $T^m{}_n$ for the generators in the antisymmetric representation. They have the following standard properties
\begin{equation}
\begin{aligned}
[t^m{}_n,t^k{}_l]&=\delta^m{}_lt^k{}_n-\delta^k{}_nt^m{}_l,\\
[T^m{}_n,T^k{}_l]&=\delta^m{}_lT^k{}_n-\delta^k{}_nT^m{}_l,\\
\PP^{k}{}_{l}{}^{m}{}_{n}\PP^{n}{}_{m}{}^{p}{}_{q}&=\PP^{k}{}_{l}{}^{p}{}_{q},\\
\PP^{mn}{}_{pq}{}^{kl}{}_{uv}\PP^{uv}{}_{kl}{}^{rs}{}_{ij}&=\PP^{mn}{}_{pq}{}^{rs}{}_{ij},\\
\PP^{m}{}_{n}{}^{pq}{}_{kl}\PP{}^{r}{}_{s}{}^{kl}{}_{pq}&=\PP^{m}{}_{n}{}^{r}{}_{s}.
\end{aligned}
\end{equation}
For the $\operatorname{SL}(5)$ EFT we shall generally prefer the use of a pair of fundamental indices $m,n = 1, \ldots, 5$ to label tensors transforming under $\bf 10$ rather than using a generalised index $M,N = 1,\ldots,10$. We do not include a factor of $1/2$ when contracting such pairs, leading to the above normalizations and factors in the expressions below.

This allows us to write the corresponding generalised Lie derivative of an object $V^m$ with conformal weight $\l$ as
\begin{equation}
{\cal L}_{\Lambda}V^m=\fr12\Lambda^{kl}\partial_{kl}V^m-\fr{\sqrt{3}}{2} \mathbb{P}^m{}_n{}^{pq}{}_{kl}
\partial_{pq}\Lambda^{kl}V^n+\fr{\lambda}{2} \partial_{kl}\Lambda^{kl}V^m.
\end{equation}
For consistency, the weight of a generalised vector $\L^{mn}$ must be set to $\l(\L)=1/5$. The generalised metric discussed below will also have weight $\l(m^{mn})=1/5$. Note that this convention is different from \cite{Musaev:2015ces} which is related to a different realization of the generalised metric in terms of the fundamental fields, and the fact that it has determinant not equal to one. The overall factor of $1/2$ is needed to make connection to the conventional $\operatorname{GL}(4)$ Lie derivative when solving the section constraint (see \cite{Berman:2011jh,Blair:2014zba}).

Since only the relative coefficients are fixed one is free to choose the factor in front of the projector and this, in turn, defines the remaining factors. We fix our coefficients to match common conventions in the literature \cite{Berman:2011jh, Blair:2013gqa,Berman:2011cg}. To ensure consistency, we impose that such transformations close onto the following algebra
\begin{equation}
\begin{aligned}
[\LL_X, \LL_Y]&=\LL_{[X,Y]_E},\\
[X,Y]_E&=\fr{1}{2}\big(\LL_X Y-\LL_Y X\big),
\end{aligned}
\end{equation}
with the following section constraint 
\begin{equation}
\dt_{[mn}\otimes \dt_{kl]}=0.
\end{equation}
Decomposing the projector into Kronecker symbols one obtains the following simple expression for the transformation of a vector of weight $\l$:
\begin{equation}
\LL_X V^{m}=\fr12{X}^{k l} {\partial}_{kl}{{V}^{m}}\,  +   {\partial}_{kn}{{X}^{m k}}\,  {V}^{n} + \fr12\Big(\lambda+\fr{2}{5}\Big)  {\partial}_{kl}{{X}^{k l}}\,  {V}^{m},
\end{equation}
which indeed coincides with \cite{Blair:2013gqa,Berman:2011cg}. For the generalised metric we have
\begin{equation}
\LL_X m^{mn}=\fr12{X}^{k l} {\partial}_{kl}{{m}^{m n}}\,  +  {\partial}_{kl}{{X}^{m k}}\,  {m}^{l n}\,  +  {\partial}_{kl}{{X}^{n k}}\,  {m}^{m l} +   \fr12{\partial}_{kl}{{X}^{k l}}m^{mn}.
\end{equation}

As an element of $\mathbb{R}^+ \times \operatorname{SL}(5) / \operatorname{SO}(5)$ the generalised metric is represented as the following matrix in terms of the fundamental fields\footnote{Note that we have decomposed what is conventionally called the generalised metric in EFT $\mathcal{M}_{MN}$ (with $M,N = 1 , \cdots, 10$) into a `small' generalised metric $m_{mn}$ by labelling the antisymmetric representation by a pair of fundamental indices $m,n = 1, \cdots, 5$ of $\operatorname{SL}(5)$. The two are related by $\mathcal{M}_{MN} = \mathcal{M}_{mn,kl} = m_{mk} m_{ln} - m_{ml} m_{kn}$.}

\begin{gather}
\begin{gathered}
\label{eq:LittleGenMetric}
m_{mn}=
e^{-\fr{\ff}{2}}\begin{bmatrix}
h^{-\fr12}h_{ab} && V_a \\ \\
V_b && h^{\fr12}(1+ V^2)
\end{bmatrix}, \qquad   
m^{mn}=
e^{\fr{\ff}{2}}\begin{bmatrix}
h^{\fr12} \left( h^{ab} + V^a V^b\right) && -V^a \\ \\
- V^b && h^{-\fr12}
\end{bmatrix},\\ \\
V^a=\fr{1}{3!}\ve^{abcd}C_{bcd}=\fr{1}{3!}h^{-1/2}\e^{abcd}C_{bcd}.
\end{gathered}
\end{gather}
Here the factor $e^{\ff}= g^{\fr17}$ is given in terms of the determinant of the seven-dimensional external metric and is needed to make the generalised metric transform as a generalised tensor (see \cite{Blair:2014zba}). This will result in a transformation of the external 7-dimensional metric $g_{\m \n}$ under a U-duality rotation of the internal space. Alternate field parametrizations, where the SL(5) transformations do not act on the external volume, are also possible but we use the one above to conform with~\cite{Musaev:2015ces,Blair:2014zba}. The corresponding vielbein and its inverse read
\begin{equation}
\begin{aligned}
V^{\bm}{}_m=
e^{-\fr{\ff}{4}}\begin{bmatrix}
e^{-\fr12}e^{\ba}{}_a && e^{\fr12}V^{\ba} \\
\\
0 && e^{\fr12}
\end{bmatrix},&&
V^m{}_{\bm}=
e^{\fr{\ff}{4}}\begin{bmatrix}
e^{\fr12}e^a{}_{\ba} && -e^{\fr12}V^{a} \\
\\
0 && e^{-\fr12}
\end{bmatrix},
\end{aligned}
\end{equation}
where $e=\det e^{\ba}{}_a = h^{\frac{1}{2}}$ is the square root of the determinant of the metric $h_{ab}$ in the internal sector. In what follows, we will also be using a different parametrization of the generalised vielbein based on a trivector $\W^{abc}$, for which we have
\begin{equation}
\begin{aligned}
V^{\bm}{}_m=
e^{-\fr{\ff}{4}}\begin{bmatrix}
e^{-\fr12}e^{\ba}{}_a && 0 \\
\\
e^{-\fr12}W_{a} && e^{\fr12}
\end{bmatrix},&&
V^m{}_{\bm}=
e^{\fr{\ff}{4}}\begin{bmatrix}
e^{\fr12}e^a{}_{\ba} && 0 \\ \\
-e^{-\fr12}W_{\ba}  && e^{-\fr12}
\end{bmatrix},
\end{aligned}
\end{equation}
with 
\begin{equation}
W_{a}=\fr{1}{3!}\ve_{abcd}\W^{bcd}.
\end{equation}
The generalised metric for this parametrisation is then given by
\begin{align}
\label{eq:NonGeomGenMetric}
m_{mn} = e^{-\frac{\phi}{2}}
\begin{bmatrix}
h^{-\frac{1}{2}} (h_{ab} + W_a W_b) & W_a\\
W_b & h^{\frac{1}{2}}
\end{bmatrix}, \qquad m^{mn} = e^{\frac{\phi}{2}}
\begin{bmatrix}
h^{\frac{1}{2}} h^{ab} & - W^a\\
- W^b & h^{-\frac{1}{2}} (1 + W^2)
\end{bmatrix}
\end{align}
As was the case for the T-duality orbit of the NS5-brane in \cite{Bakhmatov:2016kfn} it will be more natural to calculate flux components using the above parametrization.

Finally, we couple the internal sector to a seven-dimensional external space (indexed by $\mu, \nu = 1, \cdots, 7$) to write the full SL(5) EFT action as
\begin{equation}
\begin{aligned}
\LL_{\text{EFT}}=&\ \LL_{\text{EH}}(\hat R)+\LL_{\text{sc}}(\mc{D}_\m 
m_{kl})+\LL_{\text{V}}(\F_{\m\n}{}^{mn})+\LL_{\text{T}}(\F_{\m\n\r\, m})\\
&+\LL_{\text{top}}- g^{\frac{1}{2}} m^{-1}V(m_{kl},g_{\m\n}).
\end{aligned}
\end{equation}
where the kinetic terms are given by
\begin{equation}
\begin{aligned}
\LL_{\text{EH}}&= g^{\frac{1}{2}} \hat{R}= g^{\frac{1}{2}} e^\m{}_{\bar{\mu}} e^\n{}_{\bar{\nu}}
\hat{R}_{\m\n}{}^{\bar{\mu} \bar{\nu}},\\
\LL_{\text{sc}}&=\fr{1}{4} g^{\frac{1}{2}} g^{\m\n}\mc{D}_\m m_{mn}\mc{D}_\n m^{mn},\\
\LL_{\text{V}}&=-\fr{1}{4} g^{\frac{1}{2}} m_{mk}m_{nl}\F_{\m\n}{}^{mn}\F^{\m\n 
kl},\\
\LL_{\text{T}}&=-\fr{2}{3\cdot(16)^2} g^{\frac{1}{2}} m^{mn}\F_{\m\n\r\, m}\F^{\m\n\r}{}_{n},
\end{aligned}
\end{equation}
and the potential term is as follows
\begin{equation}
\begin{aligned}
-m^{-1}V&= \fr18 {m}^{m k} {m}^{n l}  {\partial}_{mn}{{m}^{p q}}\,  {\partial}_{kl}{{m}_{p q}}\,  + \fr12  {\partial}_{mn}{{m}^{m k}}\,  {\partial}_{kl}{{m}^{n l}}\, -\fr12  {m}^{m k} {m}^{n p} {\partial}_{kl}{{m}_{p q}}\,  {\partial}_{mn}{{m}^{l q}}\,  \\ &-{\partial}_{mn}{{m}^{m k}}\,  {m}^{p q} {\partial}_{kl}{{m}_{p q}}\,  + \fr58  {m}^{n l} {m}^{m k} {m}^{r s} {\partial}_{mn}{{m}_{r s}}\,  {m}^{p q} {\partial}_{kl}{{m}_{p q}} \\
&+\frac{1}{2} g^{\mu \nu}\partial_{mn}{g_{\mu \nu}}\partial_{kl}{m^{m k}}m^{n l}
+\frac{1}{8} m^{m k}m^{n l}g^{\mu \nu}\partial_{mn}{g_{\mu \nu}}g^{\rho \sigma}\partial_{kl}{g_{\rho \sigma}} \\
&+\frac{1}{8} m^{m k}m^{n l}\partial_{mn}{g^{\mu \nu}}\partial_{kl}{g_{\mu \nu}} -\frac{1}{2} m^{m k}m^{n l}g^{\mu \nu}\partial_{mn}{g_{\mu \nu}}m^{p q}\partial_{kl}{m_{p q}}
\end{aligned}
\end{equation}
with the following definitions
\begin{equation}
\begin{aligned}
\hat{R}_{\m\n \bar{\mu} \bar{\nu}}&=R_{\m\n \bar{\mu} \bar{\nu}}+\F_{\m\n}{}^{mn}e^\r{}_{\bar{\mu}}\dt_{mn} e_{\r \bar{\nu}},\\
0&=\mc{D}_{[\m} e^{\bar{\nu}}{}_{\n]}-\fr14 \w_{[\m}{}^{\bar{\mu} \bar{\nu}}e_{\n]\bar{\nu}},\\
g&=\det ||g_{\mu \nu}||,\\
m&=\det ||m_{mn}||.
\end{aligned}
\end{equation}
In addition, the variation of the Chern-Simons term is given by
\begin{equation}
\d \mc{L}_{top}= -\fr{1}{8(4!)^2}\e^{\m\n\r\l\s\t\k}\bigg[\F_{\m\n\r\l}{}^{m}\dt_{mn} \D C_{\s\t\k}^{n}
+6\F_{\m\n}{}^{mn}\F_{\r\l\s m}\D B_{\t\k n}-2\F_{\m\n\r m}\F_{\l\s\t n}\d 
A_\k^{mn}\bigg],
\end{equation}
where we have corrected a misprint in the prefactor in the same formula in \cite{Musaev:2015ces}.

The uplift of a solution of 11D supergravity, written only in terms of the metric, into a thus defined EFT is then trivial. Indeed, one just starts with a 11=7+4 split of the solution and substitutes the fields into the generalised metric. The fact that the 11D supergravity is fully embedded inside the EFT guarantees that such a background is indeed a solution. Then one is free to rotate the fields by an SL(5) transformation to recover other solutions of EFT which, however, may no longer be geometric solutions of 11D supergravity.

\subsection{KK6 monopole}

Let us now consider the case of the KK-monopole of 11D supergravity, which a coordinate arrangement described in Table \ref{tab:KK6}. Note that this fills the seven-dimensional `external' space-time completely.
\begin{table}[H]
\centering
\begin{tabulary}{\textwidth}{LCCCCCCCCCCCCCC}
\toprule
& $t$ & $x^1$ & $x^2$ & $x^3$ & $x^4$ & $x^5$ & $x^6$ & & $y^1$ & $y^2$ & $y^3$ & $y^4 = z$\\
\cmidrule{2-8}\cmidrule{10-13}
KK6 & $\times$ & $\times$ & $\times$ & $\times$ & $\times$ & $\times$ & $\times$ & & $\bullet$ & $\bullet$ & $\bullet$ & $\odot$\\
\bottomrule
\end{tabulary}
\caption{
The identification of the internal space under a $7+4$ Kaluza-Klein split of the KK6 background. Here, $\bullet$ denotes the coordinates upon which the harmonic function depends, $\times$ denotes worldvolume coordinates and $\odot$ denotes the Hopf cycle of the monopole solution.\label{tabKK6}
}
\label{tab:KK6}
\end{table} 
Under a 7+4 split, the background takes the following form:
\begin{gather}
\begin{gathered}
\textrm{d}s^2=g_{\m\n}\textrm{d}x^\m \textrm{d}x^\n+h_{ab}\textrm{d}y^a \textrm{d}y^b,\\
g_{\mu \nu} = \delta_{\mu \nu}, \qquad
||h_{ab}|| =
\begin{bmatrix}
H\d_{ij}+H^{-1}A_iA_j & H^{-1}A_i \\
H^{-1}A_j & H^{-1}
\end{bmatrix},\\
H(y)=1+\fr{h}{r}, \qquad r^2=\d_{ij}y^iy^j,
\end{gathered}
\end{gather}
where the indices span $\m,\n=0,\ldots,6; a,b=1,\cdots,4$; and $i,j=1,2,3$ such that $a =(i,z)$. 

The uplift is performed trivially by identifying the geometric coordinates $\{y^a\}$ of the spacetime with the subset $\{\hY^{a5}\}$ of the extended coordinates, under a further $\operatorname{SL}(5) \hookleftarrow \operatorname{SL}(4)$ splitting as shown in Table \ref{tab:KK6b}.
\begin{table}[H]
\centering
\begin{tabulary}{\textwidth}{LCCCCCCCCCCCCCC}
\toprule
& $t$ & $x^1$ & $x^2$ & $x^3$ & $x^4$ & $x^5$ & $x^6$ & & ${\hat{Y}}^{15}$ & ${\hat{Y}}^{25}$ & ${\hat{Y}}^{35}$ & ${\hat{Y}}^{45}$ & ${\hat{Y}}^{ab}$\\
\cmidrule{2-8}\cmidrule{10-14}
KK6 & $\times$ & $\times$ & $\times$ & $\times$ & $\times$ & $\times$ & $\times$ & & $\bullet$ & $\bullet$ & $\bullet$ & $\odot$ & \\
\bottomrule
\end{tabulary}
\caption{
The trivial lift of the KK6 background to EFT, identifying the internal space with a subset of the extended space. In this frame, the section aligns with the eleven-dimensional physical spacetime, namely $(x^\mu, {\hat{Y}}^{a5})$.
}
\label{tab:KK6b}
\end{table}

The generalised metric for the uplift of the KK6 solution is obtained by simply substituting the metric $h_{ab}$ into $m_{mn}$ in \eqref{eq:LittleGenMetric} with $C_{(3)}=0$.

\subsection{U-duality transformation of KK6-monopole}
\label{KK6_U}

To obtain the non-geometric five-brane background we consider the following $\operatorname{SL}(5)$ rotation
\begin{equation}
\label{eq:w11}
\begin{aligned}
T^m \longrightarrow U^m{}_n T^n, \qquad
U^m{}_n=
\begin{bmatrix}
1 & 0 & 0 & 0 & 0 \\ 
0 & 1 & 0 & 0 & 0 \\ 
0 & 0 & 1 & 0 & 0 \\ 
0 & 0 & 0 & 0 & 1 \\ 
0 & 0 & 0 & -1 & 0 
\end{bmatrix},
\end{aligned}
\end{equation}
under which the matrix of coordinates transforms as follows:
\begin{equation}
\begin{bmatrix}
 0 & {\hat{Y}}^{12} & {\hat{Y}}^{13} & {\hat{Y}}^{14} & {\hat{Y}}^{15} \\
 -{\hat{Y}}^{12} & 0 & {\hat{Y}}^{23} & {\hat{Y}}^{24} & {\hat{Y}}^{25} \\
 -{\hat{Y}}^{13} & -{\hat{Y}}^{23} & 0 & {\hat{Y}}^{34} & {\hat{Y}}^{35} \\
 -{\hat{Y}}^{14} & -{\hat{Y}}^{24} & -{\hat{Y}}^{34} & 0 & {\hat{Y}}^{45} \\
 -{\hat{Y}}^{15} & -{\hat{Y}}^{25} & -{\hat{Y}}^{35} & -{\hat{Y}}^{45} & 0 
\end{bmatrix} \longrightarrow
\begin{bmatrix}
  0 & {\hat{Y}}^{12} & {\hat{Y}}^{13} & {\hat{Y}}^{15} & -{\hat{Y}}^{14} \\
  -{\hat{Y}}^{12} & 0 & {\hat{Y}}^{23} & {\hat{Y}}^{25} & -{\hat{Y}}^{24} \\
  -{\hat{Y}}^{13} & -{\hat{Y}}^{23} & 0 & {\hat{Y}}^{35} & -{\hat{Y}}^{34} \\
  -{\hat{Y}}^{15} & -{\hat{Y}}^{25} & -{\hat{Y}}^{35} & 0 & {\hat{Y}}^{45} \\
  {\hat{Y}}^{14} & {\hat{Y}}^{24} & {\hat{Y}}^{34} & -{\hat{Y}}^{45} & 0 \\
\end{bmatrix}.
\end{equation}
More succinctly, we represent this as
\begin{equation}
\begin{aligned}
{\hat{Y}}^{i4} \rightarrow {\hat{Y}}^{i 5}, \qquad {\hat{Y}}^{i5} \rightarrow - {\hat{Y}}^{i4}
\end{aligned}
\end{equation}
which corresponds to the U-duality transformation $U_{123}$---an element of the SL(2) subgroup of SL(5) generated by the root $\a_{45}$, as discussed earlier.
As we explicitly check below, this rotates the KK6-monopole of table~\ref{tab:KK6b} into the $6^{(3,1)}$ solution. 

The generalised metric after the rotation is given by
\begin{align}
m_{mn} =
\begin{bmatrix}
\delta_{ij} + H^{-2} A_i A_j & 0 & H^{-2} A_i\\
0 & H & 0\\
H^{-2} A_j & 0 & H^{-2}
\end{bmatrix}
\end{align}

In the non-geometric parametrisation \eqref{eq:NonGeomGenMetric}, one obtains the following background
\begin{equation}
\begin{aligned}
h_{ab}&=\mbox{diag}[H^{-1},H^{-1},H^{-1},1], 
\\
\W^{ij4}&=-\e^{ijk}A_k, \\		
e^{\f}&=H \longrightarrow g_{\m\n}=\mbox{diag}[H, \ldots, H],
\end{aligned}
\end{equation}
where $H$ becomes a harmonic function of only the extended internal coordinates
\begin{equation}
H=1+\fr{h}{\sqrt{({\hat{Y}}^{14})^2+({\hat{Y}}^{24})^2+({\hat{Y}}^{34})^2}}.
\end{equation}
Before commenting on this na\"{i}vely strange result let us also write the background in the $C$-parametrization \eqref{eq:LittleGenMetric}
\begin{equation}
\begin{aligned}
h_{ab}&=\fr{H^{-\fr53}}{\big(H^2+A^2\big)^{\fr23}}
\begin{bmatrix}
 H^2+A_1^2 & A_1 A_2 & A_1 A_3 & 0 \\
 A_1 A_2 & H^2+A_2^2 & A_2 A_3 & 0\\
 A_1 A_3 & A_2 A_3 & H^2+A_3^2 & 0 \\
 0 & 0 & 0 & H^3
\end{bmatrix}, 
\\
C_{ij4}&=-\e_{ijk}\fr{A_k}{H^2+A^2},\\		
e^{\ff}&=H^\fr13(H^2+A^2)^{\fr13} \longrightarrow g_{\m\n}=\mbox{diag}[H^\fr13 K^{\fr13}, \ldots, H^\fr13 K^{\fr13}],
\end{aligned}
\end{equation}
where $K=H^2+A^2$. 

As mentioned above the resulting solution depends only on the extended internal coordinates that are not in the original M-theory. The reason for this is that we started with a Kaluza--Klein monopole that fills all seven external dimensions and thus has its transversal directions, on which the original solution depends, completely in the internal space. As the U-duality $U_{123}$ exchanges all the geometric coordinates $y^1$, $y^2$ and $y^3$ for non-geometric coordinates we find a fully non-geometric  solution.

\begin{table}[H]
\centering
\begin{tabulary}{\textwidth}{LCCCCCCCCCCCCCCCCC}
\toprule
& $t$ & $x^1$ & $x^2$ & $x^3$ & $x^4$ & $x^5$ & $x^6$ & & ${\hat{Y}}^{15}$ & ${\hat{Y}}^{25}$ & ${\hat{Y}}^{35}$ & ${\hat{Y}}^{45}$ & ${\hat{Y}}^{14}$& ${\hat{Y}}^{24}$& ${\hat{Y}}^{34}$& $\cdots$\\
\cmidrule{2-8}\cmidrule{10-17}
KK6 & $\times$ & $\times$ & $\times$ & $\times$ & $\times$ & $\times$ & $\times$ & & $\bullet$ & $\bullet$ & $\bullet$ & $\odot$ & \\ \\
$6^{(3,1)}$ & $\times$ & $\times$ & $\times$ & $\times$ & $\times$ & $\times$ & $\times$ & & $\otimes$ & $\otimes$ & $\otimes$ & $\odot$ & $\bullet$ & $\bullet$ & $\bullet$ & \\
\bottomrule
\end{tabulary}
\caption{
Embedding of the KK6 monopole and the $6^{(3,1)}$-brane obtained by the rotation $U_{123}$. Bullets denote the directions on which the harmonic function depends, $\times$ denote the worldvolume directions and $\otimes$ denote smeared directions. The quadratic special circle $\odot$ for the $6^{(3,1)}$-brane is along the same direction as for the KK-monopole. Three cubic directions then correspond to the three non-geometric coordinates in the harmonic function.
}
\label{tab:Orbit}
\end{table}

To understand what this solution is one considers the transformation of the mass formula using the rules from Section \ref{exotic_intro}. The M-theory circle in our conventions appears to be $Y^{15}$ and the Hopf direction $Y^{45}$ is never touched in the chain. This means that we can write (reducing over $\sharp=7$)
\begin{align}
KK6:\quad& \frac{R_1R_2R_3R_4R_5R_6R_{10}^2}{\ell_{11}^9}\\
6_3^1:\quad &\frac{R_1R_2R_3R_4R_5R_6R_{10}^2}{g_s^3 \ell_{s}^9}\\
6_3^{(2,1)}: \quad &\frac{R_1R_2R_3R_4R_5R_6R_8^3R_9^3R_{10}^2}{g_s^3 \ell_{s}^{15}}\\
6^{(3,1)}: \quad & \frac{R_1R_2R_3R_4R_5R_6R_7^3R_8^3R_9^3R_{10}^2}{ \ell_{11}^{18}}.
\end{align}
 The potential for the last brane one is $A_{11,4,3}$, whose flux cannot be defined in a straightforward in the conventional supergravity. Here we conjecture that the fluxes $Q$ and $L$ corresponding to the background should be precisely the ones sourced by the 6-brane.

In addition, one can also verify the type of the new solution by considering the transformation of the associated $E_{11}$ root. This is similar to the (smeared) brane solutions discussed in~\cite{West:2004st,Cook:2004er,Englert:2007qb} and we will say more about $E_{11}$ in Section~\ref{sec:e11}. The potential of the KK6 monopole that we are starting from is a particular component of the $A_{8,1}$ potential corresponding to the dual graviton. The $E_{11}$ root for the particular alignment of the KK6 that we are choosing (in the labelling of roots in the above references) is
\begin{align}
\alpha_{KK6} &= (1,2,3,4,5,6,7,9,6,3,3) \quad \leftrightarrow \quad\mathrm{diag}(G_{\hat{M}\hat{N}})= H^{(0,0,0,0,0,0,0,-1,1,1,1)}.
\end{align}
The transformation $U_{123}$ corresponds to the fundamental Weyl reflection $w_{11}$ and leads to
\begin{align}
\alpha_{6^{(3,1)}} &= (1,2,3,4,5,6,7,9,6,3,6) \quad \leftrightarrow\quad \mathrm{diag}(G_{\hat{M}\hat{N}})= H^{(1,1,1,1,1,1,1,0,-1,-1,-1)}.
\end{align}
The notation here is such that the exponent of the harmonic function $H$ is linearly related to the root vector through a standard change of basis to the `wall basis' (see also~\cite{Damour:2002et}). Setting $A=0$ in the original KK6 and also in the transformed solution we recognise that the powers of $H$ agree with what is stated above. The roots can be traced back to the irreducible SL(11) representation they belong to inside $E_{11}$ and this gives the identification of the transformed solution as coupling to the potential $A_{11,4,3}$ (on level $\ell=6$), which interacts with $6^{(3,1)}$.

\subsection{Generalised torsion and fluxes}
\label{fluxes}

Given the generalised Lie derivative $\LL$ the generalised torsion may be defined in the usual way
\begin{equation}
\Q(U,W)=\LL^\nabla_U W-\LL^\dt_U W,
\end{equation}
where the superscript denotes the derivative used to calculate $\LL$. To make contact with the flux formulation of EFT, one defines the covariant derivative using the so-called Weitzenb\"ock connection, that is
\begin{equation}
\begin{aligned}
\nabla_{mn}U^{kl}&=\fr12\dt_{mn}U^{kl}+\fr12 \G_{mn,pq}{}^{kl}U^{pq},\\
\G_{mn,pq}{}^{kl}&=4\G_{mn,[p}{}^{[k}\d_{q]}{}^{l]},\\
\G_{mn,p}{}^q&=V^p{}_{\bp}\dt_{mn}V^{\bp}{}_q,
\end{aligned}
\end{equation}
where $V^{\bm}{}_m$ is the generalised vielbein with barred (flat) indices referring to the tangent space. Such a connection ensures that the generalised curvature tensor vanishes and that the remainder in the commutator of covariant derivatives transforms as a tensor, coinciding with the generalised torsion as defined above. Moreover, understanding the generalised vielbein as a twist matrix of a generalised Scherk--Schwarz reduction ansatz, this torsion precisely reproduces the embedding tensor of gauged supergravity, see for example~\cite{Berman:2012uy} and references therin. Since the latter is related to internal constant fluxes in compactification schemes it is natural to understand it as a general duality-covariant definition of non-constant fluxes, or with slight abuse of terminology, field strengths.

Substituting the Weitzenb\"ock connection into the definition of the torsion, one obtains the components
\begin{equation}
\begin{aligned}
\Q_{\bk \bl, \bn}{}^{\bm}&=\fr12 \delta^{\bm}_{\bn} V^{m}{}_{[\bk|}\partial_{mn}{V^{n}{}_{|\bl]}}   + \fr12 V^{m}{}_{\bk} V^{n}{}_{\bl} V^{\bm}{}_{k}  \partial_{mn}{V^{k}{}_{\bn}}\\
& \qquad + V^{\bm}{}_{k} \partial_{mn}{V^{k}{}_{\bk}} V^{m}{}_{\bl} V^{n}{}_{\bn} + \delta^{\bm}_{\bk} V^{n}{}_{\bn} \partial_{mn}{V^{m}{}_{\bl}},
\end{aligned}
\end{equation}
where we denote the flattened derivative by $\dt_{\bm\bn}:=V^{m}{}_{\bm}V^{n}{}_{\bn}\dt_{mn}$.
This may be rewritten as
\begin{equation}
\Q_{\bar{k}\bar{l} , \bn}{}^{\bm} = \delta^{\bm }_{[\bk } Y_{\bl] \bn }  - \frac{10}{3} \delta^{\bm }_{[\bk }  \theta_{\bl] \bn }   - 2 \epsilon_{\bk\bl \bn \bp \bq } Z^{\bp \bq, \bm }  + \frac{1}{3} \theta_{\bk\bl} \delta_{\bn}^{\bm },
\end{equation}
with the irreducible components defined in the same way as in \cite{Berman:2012uy}
\begin{equation}
\label{eq:emdcomp}
\begin{aligned}
\bf{\ol{10}}: && \theta_{\bar{m} \bar{n}} & =  \frac{1}{10 } \left( V^{\bar{k}}{}_k \partial_{\bar{m} \bar{n}} V^k{}_{\bar{k}}  - V^{\bar{k}}{}_k \partial_{\bar{k} [\bar{m} } V^k{}_{\bar{n}]} \right)  \ ,\\
\bf{\ol{15}}: && Y_{\bar{m} \bar{n} }& =  V^{\bar{k}}{}_k \partial_{\bar{k} (\bar{m} } V^k{}_{\bar{n})}  \ ,\\ 
\bf{40}: && Z^{\bar{m} \bar{n}, \bar{p} } &= - \frac{1}{24} \left( \epsilon^{\bar{m} \bar{n} \bar{\imath} \bar{\jmath} \bar{k} } V^{\bar{p}}{}_t \partial_{\bar{\imath}\bar{\jmath}} V^t{}_{\bar{k}}    + V^{[\bar{m}|}{}_t \partial_{\bar{\imath}\bar{\jmath}} V^{t}{}_{\bar{k}}  \epsilon^{|\bar{n}]  \bar{\imath} \bar{\jmath} \bar{k}  \bar{p}}  \right)  .
\end{aligned}
\end{equation}
Substituting in the explicit expression for the generalised vielbein and performing the decomposition $\operatorname{SL}(5)\times \RR^+ \hookleftarrow \operatorname{GL}(4)$, one may rewrite the irreducible components of the torsion in terms of the fundamental fields. As we will be mostly focusing on only the KK-monopole and the $6^{(3,1)}$-brane, we shall not present the general expression here. The interested reader may find them in \cite{Blair:2014zba}.

Let us now turn to the explicit backgrounds, starting with the KK-monopole. The only non-vanishing components read
\begin{equation}
\label{eq:KK6Fluxes}
\begin{aligned}
\q_{5\bar{a}}&=-\fr14e^{a}{}_{\ba}H^{-2}\dt_{5a}H=\fr23 f_{\bb \bar{\imath}}{}^{\bb},\\
Y_{5\bar{a} }&=-\fr14 e^{a}{}_{\ba}H^{-2}\dt_{5a}H=\fr23 f_{\bb \bar{\imath}}{}^{\bb},\\
Z^{\ba\bb,\bc}&=-5 H^{\fr12}\e^{\ba\bb\bd\be}f_{\bd\be}{}^{\bc}-5 H^{\fr12}\e^{\bc\bd\be[\ba}f_{\bd\be}{}^{\bb]}.
\end{aligned}
\end{equation}
Hence, we see that only the geometric flux $f_{ab}{}^c$ along with its trace is turned on---precisely the fluxes associated to the KK-monopole. 

To avoid confusion, it is worth noting that it is not surprising to see the trombone gauging turned on. Indeed, the same observation has been made in \cite{Bakhmatov:2016kfn} where all the obtained backgrounds, including the KK-monopole, have been shown to trigger the $\F_{\bar{M}}$ gaugings of DFT. Similar to the trombone gauging, these should always be set to zero in consistent truncations on compact manifolds. Since we are considering the full theory in 7+(4+6) dimensions, the background obtained here is clearly not a consistent truncation into either 7 or 4 dimensions and the non-zero trombone gauging of the generalised torsion is a perfect indication of this. We thus see that using the word ``gauging'' for components of the generalised torsion is somewhat misleading and is used here only for the sake of succinctness and to avoid usage of the even more confusing ``field strength''.

Substituting the background of the full 11D KK-monopole into the expressions for the fluxes above, we obtain that the only non-vanishing components in the $\bf 20$ are 
\begin{equation}
\begin{aligned}
Z^{4[\bar{\imath},\bar{\jmath}]}&=-5H^{\fr12}e^{\bar{\imath}}{}_i e^{\bar{\jmath}}{}_j\e^{ijk}\dt_k H,\\
Z^{4\bar{\imath},4}&=15 H^{-\fr12}e^{\bar{\imath}}{}_i\dt_i H.
\end{aligned}
\end{equation}
According to the SL(5) duality orbits these components transform into the $Q_{[4,i]}\in {\bf 6}_{-2}^{(Z)}$ and $L^i \in {\bf 4}_{-7}^{(Z)}$, which correspond to the flux of the  $6^{(3,1)}$-brane as shown above. 

Now we turn to the five-brane bound state background, for which we find the following non-zero components of the generalised torsion:
\begin{equation}
\label{eq:53Fluxes}
\begin{aligned}
\bf{6}:  && \q_{\ba\bb}&=\fr14e^{4}{}_{[\ba}e^{a}{}_{\bb]}H^{-1}\dt_{a4}H=\d^4_{[\ba}Q_{\bb],4},\\
\bf{10}: && Y_{\ba\bb}&=\fr14e^{4}{}_{(\ba}e^{a}{}_{\bb)}H^{-1}\dt_{a4}H=\d^4_{(\ba}Q_{\bb), 4},\\
\bf{6_{-2}}:  && Z^{5[\ba,\bb]}&=5 \e^{\ba\bb\bc\bd}e^{4}{}_{[\bc}e^{a}{}_{\bd]}H^{-1}\dt_{a4}H=-20 \e^{\ba\bb\bc 4}Q_{\bc,4},\\
\bf{4_{-7}}: &&  Z^{5\ba,5}& =\fr{15}{2}e^{\ba}{}_a\e^{abcd}A_{b}\dt_{cd}H^{-1}=L^{\ba}
\end{aligned}
\end{equation}
Hence, we see that the only non-vanishing flux components $Q_{\ba,\bb}=e^4{}_{\ba}e^b{}_{\bb}H^{-1}\dt_{4b}H$ and  $L^{\ba}=15/2e^{\ba}{}_a\e^{abcd}A_{b}\dt_{cd}H^{-1}$. This is a clear sign that the obtained background is indeed non-geometric. Note that there is no need to check that these components satisfy the Bianchi identities as these are automatically satisfied by the generalised torsion, by construction.

\section{\texorpdfstring{$E_{11}$}{E11} embedding and U-duality orbits}
\label{sec:e11}

In this section, we revisit the U-duality transformations of the previous section by using an alternate identification of the solutions using the roots of $E_{11}$. The conventions we are using for labelling these roots is the same as in~\cite{West:2001as,Kleinschmidt:2003mf}. In particular, the exceptional root of $E_{11}$ will be labelled by root number $11$ and its associated simple Weyl reflection $w_{11}$ implements a double T-duality in the last two directions $8$ and $9$ of the ten-dimensional space-time, whilst also interchanging them. The S-duality node has number $10$ and its associated Weyl reflection $w_{10}$ implements S-duality.

As in previous work on solutions and $E_{11}$~\cite{Englert:2003zs,Englert:2003py,West:2004st,Cook:2004er,West:2004iz,Englert:2007qb,Cook:2008bi}, all solutions couple electrically to their corresponding potentials e.g.\ the M5-brane couples to a six-form potential. The potentials and U-duality orbits can then be studied in an appropriate decomposition $E_{11}\to E_{11-D}\times \operatorname{GL}(D)$~\cite{Riccioni:2007au,Bergshoeff:2007qi}. Moreover, using the identification of generalised space-time with the fundamental representation of $E_{11}$, often called $\ell_1$ in the literature~\cite{West:2003fc,Kleinschmidt:2003jf}, we can also perform the U-duality transformations directly on the coordinates. 



Recall that the coordinates $\hat{Y}^{mn}$ transform in the ${\bf 10}$ (with derivatives in the $\ol{\bf 10}$) of $\operatorname{SL}(5)$ and they decompose them in the M-theory ${\bf 4}$ solution of the section condition as
\begin{align}
\underbrace{\bf 10}_{\hat{Y}^{mn}} &\to \underbrace{\bf 4}_{y^a=\hat{Y}^{a5}} \oplus \underbrace{\bf 6}_{y^{ab}=\hat{Y}^{ab}}\,.
\end{align}
We identify the directions $y^a$ (for $a=1,2,3,4$) with the directions $x^\sharp$, $x^9$, $x^8$, $x^{7}$ (in that order) in the tables below.

\subsubsection*{From M5 to \texorpdfstring{$5^3$}{53}}

To illustrate the procedure let us consider the simple example of the non-gravitational duality chain M$5\to$NS5$\to 5_2^2\to 5^3$. This can be arranged for example as follows (see also~\cite[Sec.~5.3]{Blair:2014zba}).
\begin{table}[H]
\setlength{\tabcolsep}{3pt}
\centering
\begin{tabulary}{\textwidth}{LLCCCCCCCCCCCCLC}
\toprule
Theory & Object& 0 & 1 & 2 & 3 & 4 & 5 & 6 && 7 & 8 & 9 & $(\sharp)$ & Potential & $E_{11}$ root\\
\cmidrule{1-9}\cmidrule{11-16}
M & M5 & $\times$ &$\times$&$\times$& $\times$ &$\times$ &$\times$ &$\bullet$ && $\bullet$ &  &  & & $A_{012345} $& (1,2,3,4,5,6,6,6,4,2,2)\\
IIA & NS5 &  $\times$ &$\times$&$\times$& $\times$ &$\times$ &$\times$ &$\bullet$ && $\bullet$ &  &  & ()&  $A_{012345} $& (1,2,3,4,5,6,6,6,4,2,2)\\
$\genfrac{}{}{0pt}{}{T_8T_9}{8\leftrightarrow 9}$ &&&&&&&&&&&&&&& $\downarrow w_{11}$\\
IIA & $5_2^2$ & $\times$ &$\times$&$\times$& $\times$ &$\times$ &$\times$ &$\bullet$ && $\bullet$ & $\odot$ &$\odot$  &($\odot$) & $A_{01234589|89} $& (1,2,3,4,5,6,6,6,4,2,4)\\
M & $5^3$ & $\times$ &$\times$&$\times$& $\times$ &$\times$ &$\times$ &$\bullet$ && $\bullet$ & $\odot$ & $\odot$ &$\odot$ & $A_{01234589\sharp|89\sharp} $& (1,2,3,4,5,6,6,6,4,2,4)\\
\bottomrule
\end{tabulary}
\caption{\label{tabM553} Traditional SL(5) U-duality orbit of a smeared M5 brane to a $5^3$-brane with corresponding potentials and $E_{11}$ roots. The $\bullet$ indicates the two non-compact directions on which the smeared M5 brane depends while $\odot$ denotes directions that appear quadratically in the BPS mass formula. $\times$ are world-volume directions.}
\setlength{\tabcolsep}{6pt}
\end{table}

One can follow the masses through the various steps using the rules~\eqref{TS} and~\eqref{M2II}  to verify the various objects. Note that the M5 brane that we start with is smeared in three of its transverse directions.

Let us first try to understand this process. The original smeared M5 brane has a field strength component
\begin{align}
F_{0123456} = \partial_6 A_{012345}\,.
\end{align}
This component is a singlet of SL(4) and of SL(7). (There are also other field strength components, but they are not SL(4) singlets nor SL(7) singlets.) The Weyl transformation brings this to
\begin{align}
\partial_6 A_{01234589\sharp,89\sharp}
\end{align}
since the external coordinate $x^6$ is invariant under the Weyl transformation $w_{11}$. This is part of a field strength of type $F_{10,3}=*_{11} F^1{}_3$~\cite{Bossard:2017wxl} that decomposes into a singlet under SL(7) and the ${\bf 10}$ of SL(4) that is conjugate to the flux of the $5^3$ in the ${\bf\ol{10}}$.

We can also rephrase this in its dual version as follows. The original singlet flux becomes, using the background metric, dual to the components $F_{abcd}$ since
\begin{align}
F_{0123456} = *_{11} F_{789\sharp}
\end{align}
and thus can be thought of as deriving from the three-form potential $A_{abc}$ depending on the coordinates $y^a=\hat{Y}^{a5}$, from which we obtain the singlet flux $F_{abcd} = 4 \partial_{[a} A_{bcd]}$. At this point we can consider the M5 brane to be unsmeared.

If we perform the U-duality transformation $U_{123}$ on this configuration, the potential $A_{abc}$ is mapped to the dual non-geometric potential $\Omega^{abc}$ while the geometric coordinates $\hat{Y}^{a5}$ are mapped to the non-geometric coordinates $\hat{Y}^{a4}$ (except for $\hat{Y}^{45}=y^4=x^7$ that stays invariant). The singlet flux is mapped to a flux living in the ${\bf\ol{10}}$ as needed for the $5^3$ brane.

\subsubsection*{From KK6 to \texorpdfstring{$6^{(3,1)}$}{631}}

Let us now investigate the gravitational U-duality chain considered in Section \ref{KK6_U}, that is KK6$\to 6_3^1 \to 6_3^{(2,1)} \to 6^{(3,1)}$. Recall that in this construction the M-theory circle is in one of the transverse directions to the KK6 monopole. Maintaining the standard labelling of $E_{11}$ roots, the corresponding arrangement is as follows
\begin{table}[H]
\setlength{\tabcolsep}{2pt}
\centering
\begin{tabulary}{\textwidth}{LLCCCCCCCCCCCCLC}
\toprule
Theory & Object& 0 & 1 & 2 & 3 & 4 & 5 & 6 && 7 & 8 & 9 & $(\sharp)$ & Potential & $E_{11}$ root\\
\cmidrule{1-9}\cmidrule{11-16}
M & KK6 & $\times$ &$\times$&$\times$& $\times$ &$\times$ &$\times$ &$\times$ && $\odot$ & $\bullet$& $\bullet$ & $\bullet$ & $A_{01234567 |7} $& (1,2,3,4,5,6,7,9,6,3,3)\\
IIA & $6_3^1$ &  $\times$ &$\times$&$\times$& $\times$ &$\times$ &$\times$ &$\times$ && $\odot$ & $\bullet$ & $\bullet$ & ()&  $A_{01234567|7} $& 
 (1,2,3,4,5,6,7,9,6,3,3)\\
$\genfrac{}{}{0pt}{}{T_8T_9}{8\leftrightarrow 9}$ &&&&&&&&&&&&&&& $\downarrow w_{11}$\\
IIA & $6_3^{(2,1)}$ & $\times$ &$\times$&$\times$& $\times$ &$\times$ &$\times$ &$\times$ && $\odot$ & $\circledcirc$ &$\circledcirc$  &($\circledcirc$) & $A_{0123456789|789|89} $&
 (1,2,3,4,5,6,7,9,6,3,6)\\
M & $6^{(3,1)}$ & $\times$ &$\times$&$\times$& $\times$ &$\times$ &$\times$ &$\times$ && $\odot$ & $\circledcirc$ &$\circledcirc$  &$\circledcirc$ & $A_{0123456789\sharp|789\sharp|89\sharp} $&  (1,2,3,4,5,6,7,9,6,3,6)\\
\bottomrule
\end{tabulary}
\caption{\label{tabKK6631} Traditional SL(5) U-duality orbit of a smeared KK6 monopole to a $6^{(3,1)}$-brane with corresponding potentials and $E_{11}$ roots. The $\bullet$ indicates non-compact directions on which the full KK6 monopole depends, these are smeared in the traditional procedure. The $\odot$ and $\circledcirc$ denote directions that appear quadratically and cubically in the BPS mass formula respectively. $\times$ are world-volume directions.}
\setlength{\tabcolsep}{6pt}
\end{table}

At the level of fluxes we have explicitly shown that the full 11D background of KK6 contains fluxes in $\bf 40$, $\bf \ol{ 15}$ and $\bf \ol{10}$ of SL(5) under the split $11=7+4$. The latter two appear since the background is not smeared and all the fluxes still depend on the ``internal'' coordinates. To see how these orbits can be reproduced from gauge potentials interacting electrically with the KK6 monopole and $6^{(3,1)}$-brane one considers the linearised theory. Then the potentials appear at the level 7 of the decomposition $E_{11}\hookleftarrow$GL(7)$\times$SL(5) and are given by 7-forms of GL(7) which belong to the $\bf 5 \oplus 45 \oplus 70$ of SL(5). Here the $\bf 5$ and $\bf 45$  correspond  completely to roots of $E_{11}$ of negative and zero length squared respectively and hence are not considered in the traditional approach to U-duality orbit. 

At this point, we are facing the following conundrum. Using the above potentials as SL(5) representations together with the SL(5) representation of the extended coordinates, it is not possible to form derivatives of the potentials sourcing the space-filling branes such that the ``field strengths'' transform in the SL(5) representations $\overline{\bf 10}\oplus\overline{\bf 15}\oplus{\bf{40}}$ corresponding to the fluxes. However, at the same time we have found using the explicit expressions~\eqref{eq:emdcomp} that our explicit solutions do source these fluxes. A possible resolution to this apparent paradox is that one has to use a background generalised metric from the solution that allows to raise and lower the SL(5) indices and thus making the group theory of the tensor product between derivatives and 7-form potentials consistent.\footnote{We note that fluxes for domain walls and solutions of gauged supergravity were discussed in~\cite{Bergshoeff:2012pm}.}

Indeed, to get a contribution of the linearised gauge potential to the linearised field strength (flux) one takes a derivative $\dt^{(\bf 10)} \in \bf 10$, which yields
\begin{equation}
\begin{aligned}
&\dt^{(\bf 10)} A_{\m_1\dots \m_7}{}^{(\bf 5)} &&\longrightarrow \F^{(\bf 10)}+ \F^{(\bf 40)} \Longleftrightarrow (\q_{mn}, Z^{mn,k}),\\
&\dt^{(\bf 10)} A_{\m_1\dots \m_7}{}^{(\bf 45)} &&\longrightarrow \F^{(\bf 10)}+ \F^{(\bf 15)} \Longleftrightarrow (\q_{mn}, Y_{mn}),\\
&\dt^{(\bf 10)} A_{\m_1\dots \m_7}{}^{(\bf 70)} &&\longrightarrow \F^{(\bf 10)}+ \F^{(\bf 40)} \Longleftrightarrow (Z^{mn,k}).
\end{aligned}
\end{equation}
We stress that taking the derivative in the ${\bf 10}$ rather than its defining $\overline{\bf 10}$ requires the use of an SL(5) background metric $\ol{m}_{mn}$.

Hence, one concludes that taking the full SL(5) EFT derivative of the potential in $\bf 70$ which corresponds to a positive root gives only flux in $\bf 40$ as expected. This is the conventional piece of the full picture relevant for the fully smeared KK6 potential. Such a background is a solution of the 7D supergravity and generates only constant fluxes. 

\section{Discussion}
\label{discussion}

At this point it is useful to reflect on the nature of the {\it{non-geometric}} solutions we have constructed. Throughout the paper we have followed the convention of calling the SL(5) the duality group. Strictly speaking this is not true and is the cause for much confusion in EFT and DFT. The name remains due to historical reasons but the U-duality in string and M-theory only occurs in the presence of isometries yet EFT has a global SL(5) even when non are present. Note it is also  the SL(5,${\mathbb{R}}$) whereas the duality group is SL(5,${\mathbb{Z}}$). This distinction between the group determining the EFT and the duality group is explained in detail in \cite{Berman:2014jsa}. The duality group arises when there are isometries and so there is an ambiguity in identifying the canonical spacetime section of the extended theory.
In this paper we are not therefore carrying out SL(5) U-duality transformations in this traditional sense. This is a misnomer though it is nearly universal in the literature because it was through the duality groups that EFT were discovered/constructed.

So what is one doing in carrying out the SL(5) transformation? From the point of view of EFT it is nothing at all since the theory is SL(5) symmetric. One is simply changing the orientation of the solution with respect to the choice of spacetime section. This is entirely allowed within the EFT formalism. It is not U-duality because there is not an isometry. the result is then a solution that does not necessarily have an interpretation in terms of usual supergravity. Perhaps then such solutions should be rejected. However, in fact this is the power of DFT/EFT.  As discussed previously in this paper, such solutions, through the series of works \cite{Harvey:2005ab,Jensen:2011jna,Kimura:2013fda,Kimura:2013zva,Kimura:2014bxa,Kimura:2014aja,Berman:2014jsa,Lust:2017jox} describe the effects of world sheet/volume instantons. The canonical example was the NS5 brane where precisely this procedure reproduced the detailed gauged linear sigma model cacluation of instanton effects. DFT captures these effects and (though less well tested) EFT does too.
The effects come from in DFT/EFT through what is termed as {\it{localisation}} whereby the harmonic function in a solution is allowed to depend on the extended so called {\it{winding}} directions. One can solve the harmonic equations with appropriate delta function sources directly or generate such solutions using an SL(5) rotation as we have done here. These non-geometric solutions have dependence on the novel extended coordinates. The interesting thing is that when one then calculated the fluxes of these solutions using the standard Weitzenb\"ock formula one obtains the fluxes normally associated to the non-geometric branes. In fact, a careful reader will recall that the original confusion with R-flux in the H-f-Q-R T-duality chain was indeed that there were no isometries left by the time one produced R-flux and the result had a dependence on {\it{dual}} coordinates. This is now understood within DFT that non trivial winding mode dependence  sources R-flux. This is identical to what was seen in  \cite{Bakhmatov:2016kfn} and what is seen here in the EFT sense.
Thus what has been shown here is interpreted as showing how localisation in membrane winding mode space (i.e. having a harmonic dependence on these directions) produces what is called non-geometric flux and that we have calculated the specific flux components and labelled them by what branes would normally be a source. (This explains the unusual relation with fluxes and representations, we generate them not through {\it{duality}} but through localisation in winding space).

The localisation behaviour is actually not as unfamiliar as it might seem. A similar phenomenon already occurs for RR flux and D-branes. When there are $d$ isometries one can carry out a Spin$(d,d)$ transformation of the RR-flux to obtain the fluxes associated to the different D-branes. But because of the isometries one does not generate the most general flux associated to that brane. To do that requires one to allow the harmonic function in the solution to depend on directions where previously there was an isometry, i.e. localise the brane. In this way one  generates more RR-flux components that are not generated from a simple T-duality transformation).

\section{Conclusions}

The results obtained here are the EFT analogue of the DFT description of so called Q and R fluxes as first described in \cite{Blair:2014zba}. 
\begin{equation}
\begin{aligned}
H_{\hat{\alpha} \hat{\beta} \hat{\gamma}} && \longrightarrow && f_{\hat{\alpha} \hat{\beta}}{}^{\hat{\gamma}} && \longrightarrow && Q_{\hat{\alpha}}{}^{\hat{\beta} \hat{\gamma}} && \longrightarrow && R^{\hat{\alpha} \hat{\beta} \hat{\gamma}}
\end{aligned}
\end{equation}
Although this was originally investigated with a toy model starting with H-flux on a torus (obviously not a true string background) they have since been realized by a series of genuine solutions of the DFT equations of motion, allowing for a dependence on non-geometric coordinates \cite{Bakhmatov:2016kfn}. Since such backgrounds can no longer be solutions of the equations of motion of conventional supergravity, the dependence on winding coordinates is an essential stringy feature. Indeed, such a dependence is known to be related to worldsheet instanton corrections, which are non-perturbative corrections to the two-dimensional $\sigma$-model \cite{Harvey:2005ab,Kimura:2013zva}. 

Apart from the obvious extension to the $E_{7(7)}$ U-duality group, we aim to study the Bianchi identities which are known to be crucial for string cosmological model-building. The work \cite{Lombardo:2016swq} presents the Bianchi identities for the so-called $P$-fluxes, i.e.\ those encoded in the vector-spinor $\q_{MA}$ of DFT, which are constant. See \cite{Lee:2016qwn} for recent work on the so-called $P$-fluxes in EFT. Understanding the origin of the fluxes of the Type II theories from the full U-duality covariant generalised flux, one is able to recover coordinate-dependent fluxes which include,not only fluxes of $g_s^{-3}$ branes, but also branes with lower powers of $g_s$. This is work in progress.

The generalised flux of DFT allows one to define the magnetic charge of the corresponding brane by a straightforward generalization of the usual Yang-Mills magnetic monopole charge \cite{Blair:2015eba}
\begin{equation}
\m \propto \int \F_{MNK} \textrm{d} {\hat{X}}^M \wedge \textrm{d} {\hat{X}}^N \wedge \textrm{d} {\hat{X}}^K,
\end{equation}
which, indeed, gives the expected results. The generalization to the embedding tensor of EFT is not so simple. The main problem is that the irreducible representations of the corresponding embedding tensor do not contain scalars upon contraction with $\textrm{d}{\hat{Y}}^M\wedge \textrm{d}{\hat{Y}}^N \wedge \textrm{d}{\hat{Y}}^K$. In contrast, one has to increase the number of wedged $\textrm{d}{\hat{Y}}$'s in the integrand and project them onto a specific representation. Although straightforward, this is not a technically simple calculation and the problem of magnetic charges for exceptional non-geometric branes needs to be investigated separately.


Note, in this paper, we have used the terminology `non-geometric' to refer to configurations that cannot be written in terms of the usual supergravity fields in a way that they depend only the usual geometric coordinates. All the calculations we have performed were local in extended space. It is an interesting open question to show explicitly that the global structure of the solutions is such that they requiring transition functions that are not standard diffeomorphisms but more general elements of the EFT group, similar to what happens in~\cite{Shelton:2005cf,Hull:2006qs,Hohm:2012gk,Park:2013mpa,Berman:2014jba}. These are fascinating global questions. In  ~\cite{Papadopoulos:2014ifa,Hassler:2016srl,Howe:2016ggg,duBosque:2017dfc} these global questions were addressed and various proposals made for DFT. A clear input (especially in \cite{Hassler:2016srl,Howe:2016ggg}) was from the study of so called topological T-duality of the type studied in \cite{Bouwknegt:2003wp}. The work in this paper, although only local, gives hints at how  topological T-duaity might be generalised to a topological U-duality since the EFT solutions we describe contain fluxes which one imagines would be associated with some topological data. This is far from the scope of this paper but we hope it may inform an M-theory extension of \cite{Bouwknegt:2003wp}.

\textit{Note added}: While this manuscript was being finalised, the preprint~\cite{Lust:2017bwq} appeared that addresses related questions, in particular the orbit structure for SL(5) EFT.

\subsection*{Acknowledgements}

 The work of ETM is supported by the Alexander von Humboldt Foundation. The work of IB and ETM is in part supported by the Russian Government programme of competitive growth of Kazan Federal University. DSB is supported by the STFC grant ST/P000754/1, ``String Theory, Gauge Theory and Duality'' and AK and ETM partially by DAAD PPP grant 57316852 (XSUGRA). We acknowledge discussions with Masaki Shigemori and Gianluca Inverso. We are very grateful to the referee for pointing out an error in the first version of this paper.

\appendix
\section{Conventions}

\subsection{Indices}
Here we list the indices and their ranges.
\begin{itemize}
	\item $M,N, \cdots = 1, \cdots, \operatorname{dim} {\mathcal{R}}_n$: Generalised indices (both for DFT and EFT). For $\operatorname{SL}(5)$ EFT, $\operatorname{dim} {\mathcal{R}}_4=10$ and for $\operatorname{O}(d,d)$ DFT, $\operatorname{dim}{\mathcal{R}}_{\operatorname{O}(d,d)} = 2d$
	\item $m,n, \cdots = 1, \cdots, 5$: $\operatorname{SL}(5)$ indices. For $\operatorname{SL}(5)$ EFT, we shall generally prefer an antisymmetric pair of indices $[mn]$ over the generalised index $M$
	\item $a,b, \cdots = 1, \cdots, 4$: $\operatorname{SL}(4)$ indices. $m = (a,5)$.
	\item $\alpha, \beta, \cdots = 1,2,3$: Transverse space of the KK-monopole. $a=(\alpha, z)$
	\item $\hat{\alpha}, \hat{\beta}, \cdots = 1, \cdots, d$: $\operatorname{O}(d,d)$ DFT coordinates
	\item $\mu, \nu, \cdots  = 0, \cdots, 6$: External spacetime indices for $\operatorname{SL}(5)$ EFT
	\item $\hat{\textsf{M}}, \hat{\textsf{N}} = 0, \cdots, 10$: 11D spacetime; $\hat{\textsf{M}} = ({\textsf{M}},\sharp)$
	\item $\textsf{M}, \textsf{N}, \cdots = 0, \cdots 9$: 10D spacetime
	\item $\sharp$: M-theory circle
\end{itemize}
Indices flattened with a vielbein shall be denoted with an overbar e.g. $\bar{\mu}, \bar{a}$ etc. Note the contractions of antisymmetric pairs of indices comes with a factor of $1/2$ in $\operatorname{SL}(5)$ EFT:
\begin{align}
V^M W_M \equiv \frac{1}{2} V^{mn} W_{mn},
\end{align}
except for the Kronecker delta for which $\delta^M_M = \delta^{mn}_{mn} = 10$. Antisymmetric pairs of indices have also been ordered
\begin{align}
{\hat{Y}}^{mn}, \qquad m < n
\end{align}
\subsection{Metrics, Vielbeins and Metric Determinants}
We denote the metrics, vielbeins and determinants by the following triples:
\begin{itemize}
	\item $(g_{\mu \nu}, e^{\bar{\mu}}{}_{\mu}, g)$: External metric
	\item $(\mathcal{M}_{MN}, V^{\bar{M}}{}_{M}, \mathcal{M})$: generalised Metric (for any EFT or DFT)
	\item $(m_{mn}, V^{\bar{m}}{}_{m}, m)$: $\operatorname{SL}(5)$ EFT `little' generalised metric
	\item $(h_{ab}, e^{\bar{a}}{}_{a}, h = e^2)$: Internal metric
\end{itemize}
The $\operatorname{SL}(5)$ generalised metric $\mathcal{M}_{mn,pq}$ is given in terms of the little metric by $\mathcal{M}_{mn,pq} = m_{mp} m_{qn} - m_{mq} m_{pn}$ and has a generalised vielbein $V^{\bar{m}\bar{n}}{}_{mn} = 2 V^{\bar{m}}{}_{[m} V^{\bar{n}}{}_{n]}$.

For dualisation, we denote the Levi-Civita tensor as $\varepsilon$ and the symbol (tensor density) as $\epsilon$ according to
\begin{align}
\varepsilon_{abcd} & \coloneqq \sqrt{-g} \epsilon_{abcd}\\
\varepsilon^{abcd} & \coloneqq \frac{1}{\sqrt{-g}} \epsilon^{abcd}
\end{align}

\bibliographystyle{utphys}
\bibliography{SL5_exotic}
\end{document}

%% file: Defs.tex
\def\a{\alpha} \def\b{\beta} \def\g{\gamma} \def\d{\delta} \def\e{\epsilon}
\def\ve{\varepsilon}   \def\q{\theta}
  \def\k{\kappa} \def\l{\lambda} \def\m{\mu}
\def\n{\nu} \def\x{\xi} \def\p{\pi}  \def\r{\rho}
 \def\s{\sigma} \def\t{\tau}  \def\f{\varphi}
\def\ff{\phi}  \def\y{\psi} \def\w{\omega}

\def\G{\Gamma} 
\def\D{\Delta} 
   \def\Q{\Theta}
   \def\L{\Lambda}

\def\F{\Phi}   \def\W{\Omega}

\def\ba{\bar{a}}\def\bb{{\bar{b}}}\def\bc{\bar{c}}\def\bd{\bar{d}}\def\be{\bar{e}}
\def\bk{\bar{k}}\def\bl{\bar{l}}\def\bm{\bar{m}}\def\bn{\bar{n}}\def\bp{\bar{p}}\def\bq{\bar{q}}

\def\fr{\frac}  \def\dt{\partial}

\def\qquad{\ph{=\ }}

\def\ph{\phantom}
\def\mT{\mathcal{T}}
\def\mc{\mathcal}

\def\mL{\mathcal{L}}

\def\mM{\mathcal{M}}

\def\hY{\hat{Y}}

\def\RR{\mathbb{R}}
\def\TT{\mathbb{T}}

\newcommand{\ol}[1]{\overline{#1}}
\newcommand\bqa {\begin{eqnarray}}
\newcommand\eqa {\end{eqnarray}}

\newcommand{\bear}{\begin{array}}
\newcommand{\enar}{\end{array}}

\def\beq{\begin{equation}}
\def\eeq{\end{equation}}
\def\bea{\begin{eqnarray}}
\def\eea{\end{eqnarray}}

\def\F{{\mathcal{F}}}

\def\PP{\mathbb{P}}
\def\LL{\mathcal{L}}